\documentclass[twocolumn]{aastex631}

\usepackage{graphicx}
\usepackage{upgreek}
\usepackage{url}
\usepackage{hyperref}
\usepackage{lineno}
\usepackage{ragged2e}
\usepackage{lineno}
\usepackage[T1]{fontenc}
\newcommand{\Msun}{M$_\odot$}

\newcommand{\Zsun}{Z$_\odot$}

\newcommand{\nifs}{$^{56}$Ni}
\newcommand{\kms}{\,km\,s$^{-1}$} % kilometres per second
\newcommand{\ergs}{\,erg\,s$^{-1}$} % erg per second
\newcommand{\naid}{{Na\,{\sc i D}}}

\newcommand{\css}{CSS161010}
 
\newcommand{\ha}{H$\alpha$}
\newcommand{\hb}{H$\beta$}

\newcommand{\orcid}[1]{\href{https://orcid.org/#1}{\includegraphics[width=9pt]{orcid.png}}}
\newcommand{\cow}{AT~2018cow}
\newcommand{\mrf}{AT~2020mrf}
\newcommand{\xnd}{AT~2020xnd}
\newcommand{\zr}{AT~2018zr}
\newcommand{\wey}{AT~2020wey}
\newcommand{\neh}{AT~2020neh}
\newcommand{\hei}{{He\,{\sc i}}}
\newcommand{\heii}{{He\,{\sc ii}}}

\definecolor{forestgreen}{rgb}{0.0, 0.7, 0.1}

\defcitealias{Coppejans20}{C20}
\definecolor{forestgreen}{rgb}{0.0, 0.7, 0.1}

\shorttitle{CSS161010: an LFBOT with broad blueshifted H lines}
\shortauthors{Guti\'errez et al.}
\begin{document}

\title{CSS161010: a luminous, fast blue optical transient with broad blueshifted hydrogen lines}

\author[0000-0003-2375-2064]{Claudia P. Guti\'errez}\thanks{E-mail: cgutierrez@ice.csic.es}
\affiliation{Institut d'Estudis Espacials de Catalunya (IEEC), Edifici RDIT, Campus UPC, 08860 Castelldefels (Barcelona), Spain}
\affiliation{Institute of Space Sciences (ICE, CSIC), Campus UAB, Carrer de Can Magrans, s/n, E-08193 Barcelona, Spain}
\affiliation{Finnish Centre for Astronomy with ESO (FINCA), FI-20014 University of Turku, Finland}
\affiliation{Department of Physics and Astronomy, FI-20014 University of Turku, Finland}
\author[0000-0001-7497-2994]{Seppo Mattila}
\affiliation{Department of Physics and Astronomy, FI-20014 University of Turku, Finland}
\affiliation{School of Sciences, European University Cyprus, Diogenes Street, Engomi, 1516 Nicosia, Cyprus}
\author[0000-0002-3664-8082]{Peter Lundqvist}
\affiliation{The Oskar Klein Centre, Department of Astronomy, Stockholm University, AlbaNova, SE-10691, Stockholm, Sweden}
\author[0000-0003-0599-8407]{Luc Dessart}
\affiliation{Institut d’Astrophysique de Paris, CNRS-Sorbonne Université, 98 bis boulevard Arago, F-75014 Paris, France}
\author[0000-0001-9541-0317]{Santiago Gonz\'alez-Gait\'an}
\affiliation{CENTRA, Instituto Superior T\'ecnico, Universidade de Lisboa, Av. Rovisco Pais 1, 1049-001 Lisboa, Portugal}
\author{Peter G.~Jonker}
\affiliation{Department of Astrophysics/IMAPP, Radboud University, P.O. Box 9010, 6500 GL Nijmegen, The Netherlands}
\affiliation{SRON, Netherlands Institute for Space Research, Niels Bohrweg 4, 2333 CA, Leiden, The Netherlands}
\author[0000-0002-1027-0990]{Subo Dong}
\affiliation{Department of Astronomy, School of Physics, Peking University, 5 Yiheyuan Road, Haidian District, Beijing 100871, People's Republic of China}
\affiliation{Kavli Institute of Astronomy and Astrophysics, Peking University, 5 Yiheyuan Road, Haidian District, Beijing 100871, People's Republic of China}
\affiliation{National Astronomical Observatories, Chinese Academy of Science, 20A Datun Road, Chaoyang District, Beijing 100101, People's Republic of China}
\author{Deanne Coppejans}
\affiliation{Department of Physics, University of Warwick, Gibbet Hill Road, Coventry CV4 7AL, UK }
\author{Ping Chen}
\affiliation{Department of Astronomy, School of Physics, Peking University, 5 Yiheyuan Road, Haidian District, Beijing 100871, People's Republic of China}
\affiliation{Department of Particle Physics and Astrophysics, Weizmann Institute of Science, 234 Herzl St, 76100 Rehovot, Israel}
\author[0000-0002-0326-6715]{Panos Charalampopoulos}
\affiliation{Department of Physics and Astronomy, FI-20014 University of Turku, Finland}
\author[0000-0002-1381-9125]{Nancy Elias-Rosa}
\affiliation{INAF – Osservatorio Astronomico di Padova, Vicolo dell’Osservatorio 5, Padova I-35122, Italy}
\affiliation{Institute of Space Sciences (ICE, CSIC), Campus UAB, Carrer de Can Magrans, s/n, E-08193 Barcelona, Spain}
\author[0000-0002-1022-6463]{Thomas M. Reynolds}
\affiliation{Cosmic Dawn Center (DAWN), Niels Bohr Institute, University of Copenhagen, 2200, Denmark}
\affiliation{Department of Physics and Astronomy, FI-20014 University of Turku, Finland}
\author{Christopher Kochanek}
\affiliation{Department of Astronomy, The Ohio State University, 140 West 18th Avenue, Columbus OH 43210}
\affiliation{Center for Cosmology and AstroParticle Physics, The Ohio State University, 191 W. Woodruff Avenue, Columbus OH 43210 }
\author[0000-0003-2191-1674]{Morgan Fraser}
\affiliation{UCD School of Physics, L.M.I. Main Building, Beech Hill Road, Dublin 4, D04 P7W1, Ireland}
\author[0000-0002-7259-4624]{Andrea Pastorello}
\affiliation{INAF – Osservatorio Astronomico di Padova, Vicolo dell’Osservatorio 5, Padova I-35122, Italy}
\author[0000-0002-1650-1518]{Mariusz Gromadzki}
\affiliation{Astronomical Observatory, University of Warsaw, Al. Ujazdowskie 4, 00-478 Warszawa, Poland}
\author{Jack Neustadt}
\affiliation{Department of Astronomy, The Ohio State University, 140 West 18th Avenue, Columbus OH 43210}
\affiliation{Center for Cosmology and AstroParticle Physics, The Ohio State University, 191 W. Woodruff Avenue, Columbus OH 43210 }
\author[0000-0002-3256-0016]{Stefano Benetti}
\affiliation{INAF – Osservatorio Astronomico di Padova, Vicolo dell’Osservatorio 5, Padova I-35122, Italy}
\author[0000-0001-8257-3512]{Erkki Kankare}
\affiliation{Department of Physics and Astronomy, FI-20014 University of Turku, Finland}
\author[0000-0002-5477-0217]{Tuomas Kangas}
\affiliation{Finnish Centre for Astronomy with ESO (FINCA), FI-20014 University of Turku, Finland}
\affiliation{Department of Physics and Astronomy, FI-20014 University of Turku, Finland}
\author[0000-0001-5455-3653]{Rubina Kotak}
\affiliation{Department of Physics and Astronomy, FI-20014 University of Turku, Finland}
\author[0000-0002-5571-1833]{Maximilian D. Stritzinger}
\affiliation{Department of Physics and Astronomy, Aarhus University, Ny Munkegade 120, DK-8000 Aarhus C, Denmark}
\author{Thomas Wevers}
\affiliation{Space Telescope Science Institute, 3700 San Martin Drive, Baltimore, MD 21218, USA}
\affiliation{European Southern Observatory, Alonso de Córdova 3107, Vitacura, Santiago, Chile}
\author[0000-0002-9725-2524]{Bing Zhang}
\affiliation{Nevada Center for Astrophysics, University of Nevada, Las Vegas, 4505 South Maryland Parkway, Las Vegas, NV 89154, USA}
\affiliation{Department of Physics and Astronomy, University of Nevada, Las Vegas, 4505 South Maryland Parkway, Las Vegas, NV 89154, USA}
\author{David Bersier}
\affiliation{Astrophysics Research Institute, Liverpool Science Park IC2, Liverpool L3 5RF}
\author{Subhash Bose}
\affiliation{Department of Astronomy, The Ohio State University, 140 West 18th Avenue, Columbus OH 43210}
\affiliation{Department of Physics and Astronomy, Aarhus University, Ny Munkegade 120, DK-8000 Aarhus C, Denmark}
\author{David A. H. Buckley}
\affiliation{South African Astronomical Observatory, PO Box 9, Observatory, 7935 Cape Town, South Africa}
\affiliation{Department of Astronomy, University of Cape Town, Private Bag X3, 7701 Rondebosch, South Africa}
\affiliation{Department of Physics, University of the Free State, PO Box 339,9300 Bloemfontein, South Africa}
\author{Raya Dastidar}
\affiliation{Millennium Institute of Astrophysics MAS, Nuncio Monse\~nor Sotero Sanz 100, Providencia, Santiago, 8320000, Chile}
\affiliation{Instituto de Astrofisica, Universidad Andres Bello, Fernandez Concha 700, Las Condes, Santiago, Chile}
\author{Anjasha Gangopadhyay}
\affiliation{Oskar Klein Centre, Department of Astronomy, Stockholm University, Albanova University Centre, SE-106 91 Stockholm, Sweden}
\author{Aleksandra Hamanowicz}
\affiliation{Space Telescope Science Institute, 3700 San Martin Drive, Baltimore, MD 21218, USA}
\author{Juna A. Kollmeier}
\affiliation{The Observatories of the Carnegie Institution for Science, 813 Santa Barbara Street, Pasadena, CA 91101, USA}
\affiliation{Canadian Institute for Theoretical Astrophysics, University of Toronto, Toronto, ON, M5S-98H, Canada}
\author{Jirong Mao}
\affiliation{Yunnan Observatories, Chinese Academy of Sciences, 650011 Kunming, Yunnan Province, People’s Republic of China}
\author{Kuntal Misra}
\affiliation{Aryabhatta Research Institute of Observational Sciences, Nainital-263001, India}
\author{Stephen. B. Potter}
\affiliation{South African Astronomical Observatory, PO Box 9, Observatory, 7935 Cape Town, South Africa}
\affiliation{Department of Physics, University of Johannesburg, PO Box 524, 2006 Auckland Park, South Africa}
\author{Jose L. Prieto}
\affiliation{Instituto de Estudios Astrof\'isicos, Facultad de Ingenier\'ia y Ciencias, Universidad Diego Portales, Av. Ej\'ercito Libertador 441, Santiago, Chile}
\affiliation{Millennium Institute of Astrophysics MAS, Nuncio Monse\~nor Sotero Sanz 100, Providencia, Santiago, 8320000, Chile}
\author{Encarni Romero-Colmenero}
\affiliation{South African Astronomical Observatory, PO Box 9, Observatory, 7935 Cape Town, South Africa}
\affiliation{Southern African Large Telescope (SALT), P.O. Box 9, Observatory, 7935 Cape Town, South Africa}
\author{Mridweeka Singh}
\affiliation{Indian Institute of Astrophysics, Koramangala 2nd Block, Bangalore 560034, India}
\author[0000-0001-6566-9192]{Auni Somero}
\affiliation{Department of Physics and Astronomy, FI-20014 University of Turku, Finland}
\author{Giacomo Terreran}
\affiliation{Las Cumbres Observatory, 6740 Cortona Drive, Suite 102, Goleta, CA 93117-5575, USA}
\affiliation{Department of Physics, University of California, Santa Barbara, CA 93106-9530, USA}
\author{Petri Vaisanen}
\affiliation{Finnish Centre for Astronomy with ESO (FINCA), FI-20014 University of Turku, Finland}
\affiliation{South African Astronomical Observatory, PO Box 9, Observatory, 7935 Cape Town, South Africa}
\author{{\L}ukasz Wyrzykowski}
\affiliation{Astronomical Observatory, University of Warsaw, Al. Ujazdowskie 4, 00-478 Warszawa, Poland}



\begin{abstract}

We present ultraviolet, optical and near-infrared photometric and optical spectroscopic observations of the luminous, fast blue optical transient (LFBOT), CSS161010:045834-081803 (\css). The transient was found in a low-redshift ($z=0.033$) dwarf galaxy. The light curves of \css\ are characterized by an extremely fast evolution and blue colours. The $V-$band light curve shows that \css\ reaches an absolute peak of M$_{V}^{max}=-20.66\pm0.06$ mag in 3.8 days from the start of the outburst. After maximum, \css\ follows a power-law decline $\propto t^{-2.8\pm0.1}$ in all optical bands. These photometric properties are comparable to those of well-observed LFBOTs such as \cow, \mrf\ and \xnd. However, unlike these objects, the spectra of \css\ show a remarkable transformation from a blue and featureless continuum to spectra dominated by very broad, entirely blueshifted hydrogen emission lines with velocities of up to 10\% of the speed of light. The persistent blueshifted emission and the lack of any emission at the rest wavelength of \css\ are a unique feature not seen in any transient before \css. The combined observational properties of \css\ and its M$_{*}\sim10^{8}$ \Msun\ dwarf galaxy host favour the tidal disruption of a star by an intermediate-mass black hole as its origin. 
\end{abstract}



\section{Introduction} \label{sec:intro}

High cadence, wide-field sky surveys have revealed a significant number of new extra-galactic transients that show a large diversity in their spectral and photometric behaviour. Among the new types of objects, one group attracts attention due to their remarkably rapid evolution: the fast blue optical transients \citep[FBOTs;][]{Drout14, Inserra19, Pursiainen18} or fast-evolving luminous transients \citep[FELTs;][]{Rest18}. FBOTs are characterised by rise times of $<10$ days, peak absolute magnitudes of $-15>M_g^{max}>-22$ mag, and blue colours. 
Their hosts are generally found to be low-mass star-forming galaxies \citep{Drout14, Pursiainen18}. Due to their fast evolution, they are difficult to explain as supernova (SN) powered by the radioactive decay of \nifs\ \citep{Drout14}. Multiple different scenarios have been proposed to explain their properties, including shock breakout emission within a dense, surrounding wind or shell \citep[e.g.][]{Ofek10, Drout14, Rest18}, cooling envelope emission from the explosion of a star with a low-mass extended envelope with very little radioactive material \citep{Drout14}, a common envelope jet \citep{Soker19, Soker22}, a tidal disruption event (TDE) caused by an intermediate-mass black hole \citep[IMBH;][]{Perley19}, and fallback accretion \citep{Margutti19}.

Most FBOTs have been found in archival data from large imaging surveys, such as Pan-STARRS1 (PS1; \citealt{Drout14}); the Palomar Transient Factory (PTF), the Supernova Legacy Survey (SNLS; \citealt{Arcavi16}) and the Dark Energy Survey (DES; \citealt{Pursiainen18, Wiseman20}) as well as from observations by the Kepler space telescope \citep{Rest18}. However, this picture is changing thanks to surveys that monitor the sky with a cadence of a few days (e.g. the All-Sky Automated Survey for SuperNovae -- ASAS-SN, \citealt{Shappee14}; the Asteroid Terrestrial-impact Last Alert System -- ATLAS, \citealt{Tonry18, Smith20}; and the Zwicky Transient Facility -- ZTF, \citealt{Bellm19, Graham19}) and new efforts focus on finding and characterising these events in almost real-time  \citep[e.g.][]{Ho21, Perley21}.

After the discovery of \cow\ \citep{Prentice18, Perley19, Margutti19, Ho19, Chen23, Inkenhaag23}, a new luminous subclass of FBOTs, now known as LFBOTs, was identified. This class includes AT~2018lug (ZTF18abvkwla \citealt{Ho20}), CRTS-CSS161010 J045834-081803 (\css\ \citealt[hereafter C20]{Coppejans20}), \xnd\ \citep{Perley21, Ho22, Bright22}, \mrf\ \citep{Yao22}, AT~2022tsd \citep{Matthews23, Ho23} and AT~2023fhn \citep{Chrimes24, Chrimes24a}. Most of these objects have been detected in the optical, radio, and X-rays. The only exception was AT~2018lug, which had no X-ray observations. Unlike in optical photometry, where all the objects follow a similar evolution, their behaviour is more diverse in the radio and X-rays: AT~2018lug is the most luminous of these events in the radio, while AT~2022tsd is the most luminous in the X-rays. Unfortunately, due to their fast evolution, multi-epoch spectral coverage in the optical has been scarce and consists of featureless spectra except for \cow.  Therefore, despite extensive observations in different wavelengths, their nature is debated, and their origin is still unknown. However, recent observations of \cow\ at late epochs suggest that a central engine \citep{Pasham21} in the form of a black hole (BH) must be present \citep{Inkenhaag23}, although it is unclear whether it is a stellar-mass BH or an intermediate-mass BH (IMBH; \citealt{Migliori24}).

Although these LFBOTs were first detected, identified, and analysed in the optical, \css\ has only been characterised in the radio and X-rays \citepalias{Coppejans20}. We present the first ultraviolet (UV), optical and near-infrared (NIR) observations of \css. The remarkable spectral and photometric coverage allows us to study its properties in detail. In particular, the detection of broad, entirely blueshifted hydrogen lines and the available information in the X-ray and radio \citepalias{Coppejans20} may help us to provide important insights into these fast events.

The paper is organised as follows. A brief description of the observations is presented in Section~\ref{sec:obs}. Section~\ref{sec:analysis} characterises and discusses the nature of \css. In Section~\ref{sec:discus}, we discuss the origin of \css, while our summary is given in Section~\ref{sec:sum}. Throughout this work, we assume a flat $\Lambda$CDM universe, with a Hubble constant of $H_0=70$\,km\,s$^{-1}$\,Mpc$^{-1}$, and $\Omega_\mathrm{m}=0.3$.

\section{Observations of CSS161010}
\label{sec:obs}

\css\ ($\textrm{RA}=04^{\rm h}58^{\rm m}34.\!\!^{\rm{s}}41$ $\textrm{Dec}=-08^{\circ}18'03.\!\!''5$, J2000) was discovered by the Catalina Real-Time Transient Survey (CRTS; \citealt{Drake09}) on 2016 October 10 (JD = 2457671.98) at an unfiltered apparent magnitude of $16.29$. An earlier detection was obtained by ASAS-SN \citep{Shappee14} on JD $= 2457671.70$ at an apparent V-band magnitude of $16.51\pm0.12$ mag. The last non-detection obtained by ASAS-SN was on 2016 October 6 (JD$=2457667.78$; detection limit of m$_V\sim17.54$ mag). ATLAS obtained a deeper and latest non-detection on 2016 October 6 (JD$=2457668.14$; detection limit of m$_c\sim19.59$ mag). We adopt the mid-point between ATLAS's last non-detection and ASAS-SN's first detection as the start of the outburst epoch, JD$= 2457669.92\pm2.00$. 

\css\ was spectroscopically observed on 2016 October 18 \citep{CSSAtel} and reported as a blue and nearly featureless object. Given its nuclear location, fast photometric evolution and blue featureless spectrum, we started our follow-up on this date ($\sim9$ days from the start of the outburst). A total of 12 epochs of optical spectroscopy were obtained from 9.4 to 106.0 days with five different instruments, while multi-wavelength photometric coverage was obtained between 1.72 and 77.91 days. The observations and data reduction details are presented in Appendix~\ref{ap:obs}.

\section{Analysis of \css}
\label{sec:analysis}

\subsection{Galaxy}
\label{sec:gal}

\begin{figure*}
\centering
\includegraphics[width=0.92\textwidth]{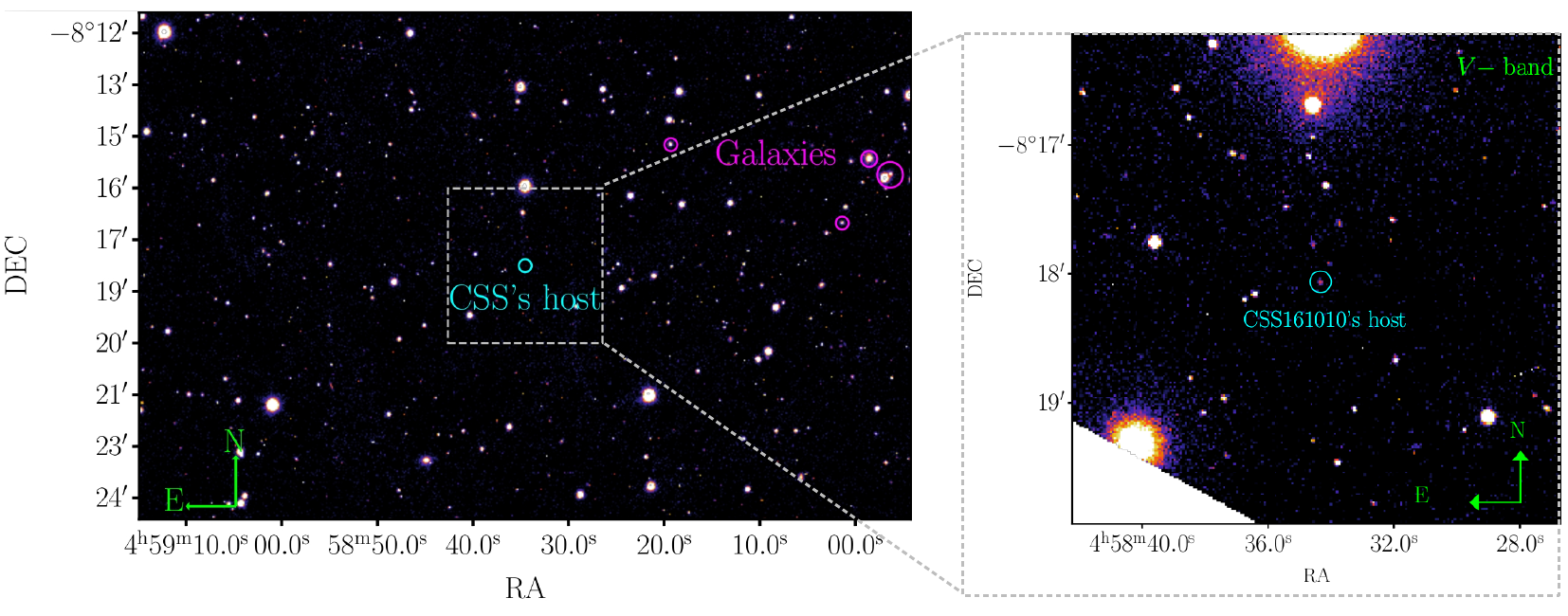}
\includegraphics[width=0.7\textwidth]{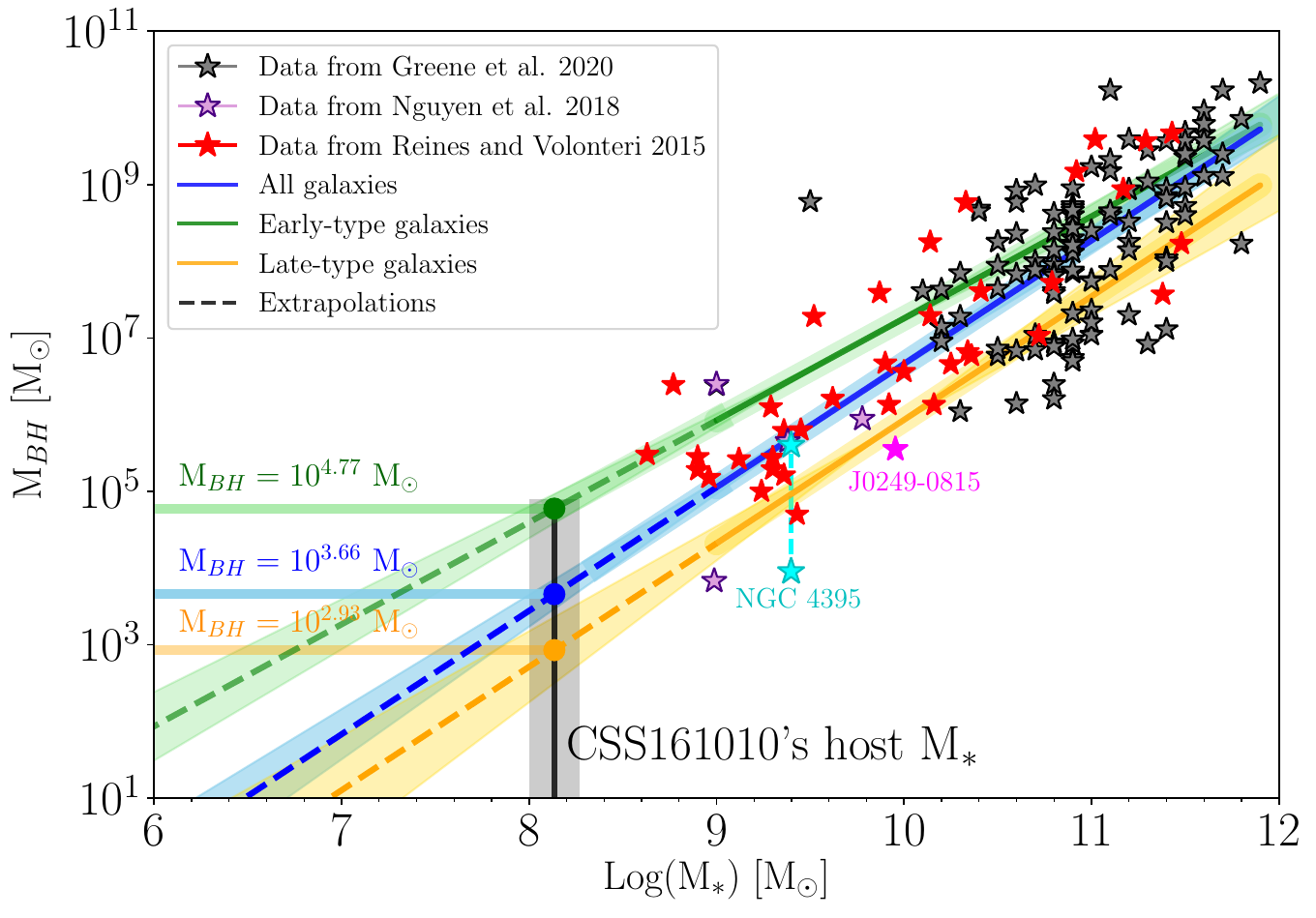}
\caption{\textbf{Top left:} PS1 (\url{https://catalogs.mast.stsci.edu/panstarrs/}) RGB false-colour $gri$ image of \css's field. The grey square is a zoom-in on the field around \css. Known galaxies in the field are marked with magenta circles. \textbf{Top right:} False-colour IMACS/Magellan V-band image of the field around \css. The transient is located close to the centre of its faint host. \textbf{Bottom:} Relationship between black hole mass (M$_{BH}$) and stellar mass (M$_{\star}$) for a sample of galaxies \citep{Reines15, Nguyen18, Greene20}, NGC~4395 (lower value, \citealt{denBrok15}; higher value, \citealt{Woo19}), J0249-0815 \citep{Zuo24}. The linear fits obtained by \cite{Greene20} for the early-type (green), late-type  (orange) and all galaxies (blue) are shown with solid lines. The extrapolations of these fits are shown with dashed lines. The estimated M$_{\star}$ of \css's host is indicated with a vertical black line, and the points where the fits cross it are highlighted with circles. 
}
\label{fig:hostgal}
\end{figure*}

The host galaxy of \css\ is WISEA J045834.37-081804.4 (Figure~\ref{fig:hostgal}), a dwarf galaxy ($M_V = -14.7\,{\rm mag}$) at $z=0.0340\pm0.0006$ (i.e., luminosity distance $D_L = 149.4$\,Mpc or a distance modulus of $\mu=35.87$). The redshift is derived from narrow emission lines visible in the transient spectra (\ha\ and [O~III] 5007 \AA). The optical spectrum of the host galaxy taken after the transient had faded below the detection threshold also confirmed this redshift (Appendix~\ref{ap:host}). Galactic reddening in the direction of \css\ is $E(B-V)=0.084$ mag \citep{Schlafly11}. Based on the absence of \naid\ absorption lines in the transient spectra and the low luminosity of the host galaxy, we assume that the host galaxy extinction towards \css\ is negligible. 
Using \textsc{prospector} \citep{Johnson21-prosp}, we find that the host's spectra and photometry are consistent with a stellar mass of log(M$_{*}/$\Msun)$=8.135^{+0.087}_{-0.079}$ (details are discussed in Appendix~\ref{ap:host}), higher than previously estimated by \citetalias{Coppejans20}. This is mostly due to the $z$-band photometry that further constrains the stellar mass.
By extrapolating the BH mass -- stellar mass correlations \citep{Greene20}, the stellar mass of the host galaxy suggests the existence of a possible BH with a mass of $10^{2.9}$ -- $10^{4.8}$ \Msun\ (Figure~\ref{fig:hostgal}), corresponding to an IMBH ($10^2$ \Msun $<M_{BH}<10^5$ \Msun; \citealt{Greene20}). While the scaling relations are firmly established for super-massive BHs, they remain poorly constrained for IMBHs, and additional observations at lower BH masses are needed to check the validity of these relations. However, IMBH mass estimates in dwarf galaxies tend to follow the extrapolation of the scaling relations into the low-mass regime \cite[e.g.][]{Reines15, Greene20}, but with a larger scatter.

\subsection{Light curves and colours}
\label{sec:LCs}

\begin{figure*}
\centering
\includegraphics[width=0.6\textwidth]{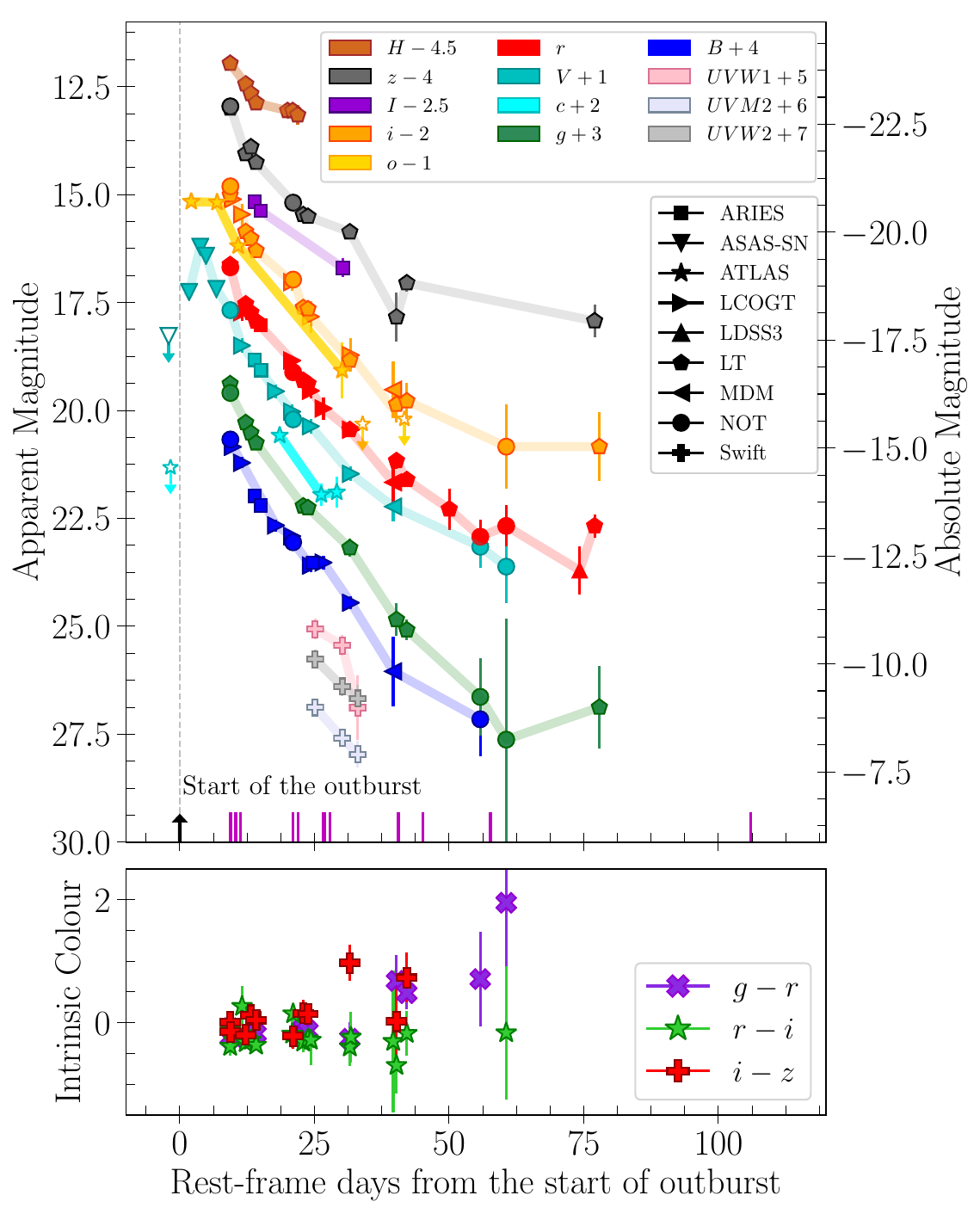}
\caption{Light and colour curves of \css. 
\textbf{Top:} UV and optical light curves of \css. Upper limits are presented as open symbols. The start of the outburst is indicated by a vertical black arrow. The vertical magenta lines are the epochs of optical spectroscopy. The photometry is host subtracted (except in the I-band) and corrected for Milky Way extinction. \textbf{Bottom:} Intrinsic colour curves of \css. 
}
\label{fig:css_LC}
\end{figure*}

In Figure~\ref{fig:css_LC} we present the light curves of \css. During the first six days after the start of the outburst, we obtained ASAS-SN V-band photometry, which allowed us to estimate a rise time of $\sim3.8$ days and a peak absolute V-band magnitude of M$_{V}^{max}=-20.66\pm0.07$ mag. In $6.3$ days from the start of the outburst (2.5 days from the maximum), \css\ declines to half its peak flux. After the peak, it follows a power-law decline of $\propto t^{-2.8\pm0.1}$ at all optical bands. Overall, the transient shows an extremely fast evolution and blue colours. These properties resemble those found in well-observed LFBOTs like \cow\ \citep{Prentice18, Perley19}, \mrf\ \citep{Yao22} and \xnd\ \citep{Perley21, Ho22}.

The intrinsic colour curves of \css\ are presented in the bottom panel of Figure~\ref{fig:css_LC}. Colour information is only available from $\sim9.4$ days post-outburst. At this point, \css\ shows blue colours ($g-r=-0.24$ mag, $r-i=-0.38$ mag), which last for $\sim12-15$ more days. From day 25 onwards, the $g-r$ and $r-i$ colours become bluer, while $i-z$ becomes redder, going from 0.14 mag at 23.8 days to 0.97 mag at 31.6 days. After this epoch, $g-r$ also becomes redder. During the full period of observations, the $r-i$ colour remains blue, with a quasi-constant evolution at a mean value of $\approx-0.24$ mag. From the imaging polarimetry obtained at $\sim53$ days from the start of the outburst, we found that \css\ shows a similar level of polarisation as the surrounding field stars (5–10\% linear polarisation). We, therefore, find that CSS161010 does not show significant polarisation above this level.

We constructed the bolometric light curve and estimated the blackbody temperatures and radii for \css\ employing the \textsc{superbol} code \citep{Nicholl18}. We used the extinction-corrected $BgVriz$ photometry. To have similar coverage in different bands at each epoch, we either interpolated or extrapolated the light curves using a low-order polynomial or obtained the magnitude from the nearest epochs using the V-band as a reference filter and assuming a constant colour. We then converted all magnitudes into fluxes at the effective wavelength of each filter and integrated them over the spectral energy distribution (SED). The flux outside the observed passbands was estimated by extrapolating the blackbody fit over all wavelengths. 

We found that \css\ reached a peak luminosity of ${\rm L}_{bol} = (1.30\pm 0.56) \times 10^{44}$ erg s$^{-1}$ at $\sim3.8$ days from the start of the outburst. After the peak, the bolometric light curve followed a power-law decline similar to that measured in the optical bands. Integrating over the observed epochs, we find a total radiated energy of $(6.62\pm0.02)\times10^{49}$ erg. From $\sim2$ to 60 days, the blackbody temperature (${\rm T}_{BB}$) shows roughly constant evolution at around ${\rm T}_{BB}\approx15000-16000$ K, while the blackbody radius (${\rm R}_{BB}$) mimics the light curve evolution: a fast rise to the peak, followed by a fast decline. It rises from R$_{BB}=1.1\times10^{15}$ cm at $\sim2$ days to ${\rm R}_{BB}=1.9\times10^{15}$ cm at the peak. After that, the radius declines continuously until it reaches a value of ${\rm R}_{BB}=3.7\times10^{13}$ cm at $\sim61$ days (see Section~\ref{sec:fbots}).

\subsection{Spectroscopic evolution}
\label{sec:Spec_evol}

\begin{figure*}
\centering
\includegraphics[width=0.6\textwidth]{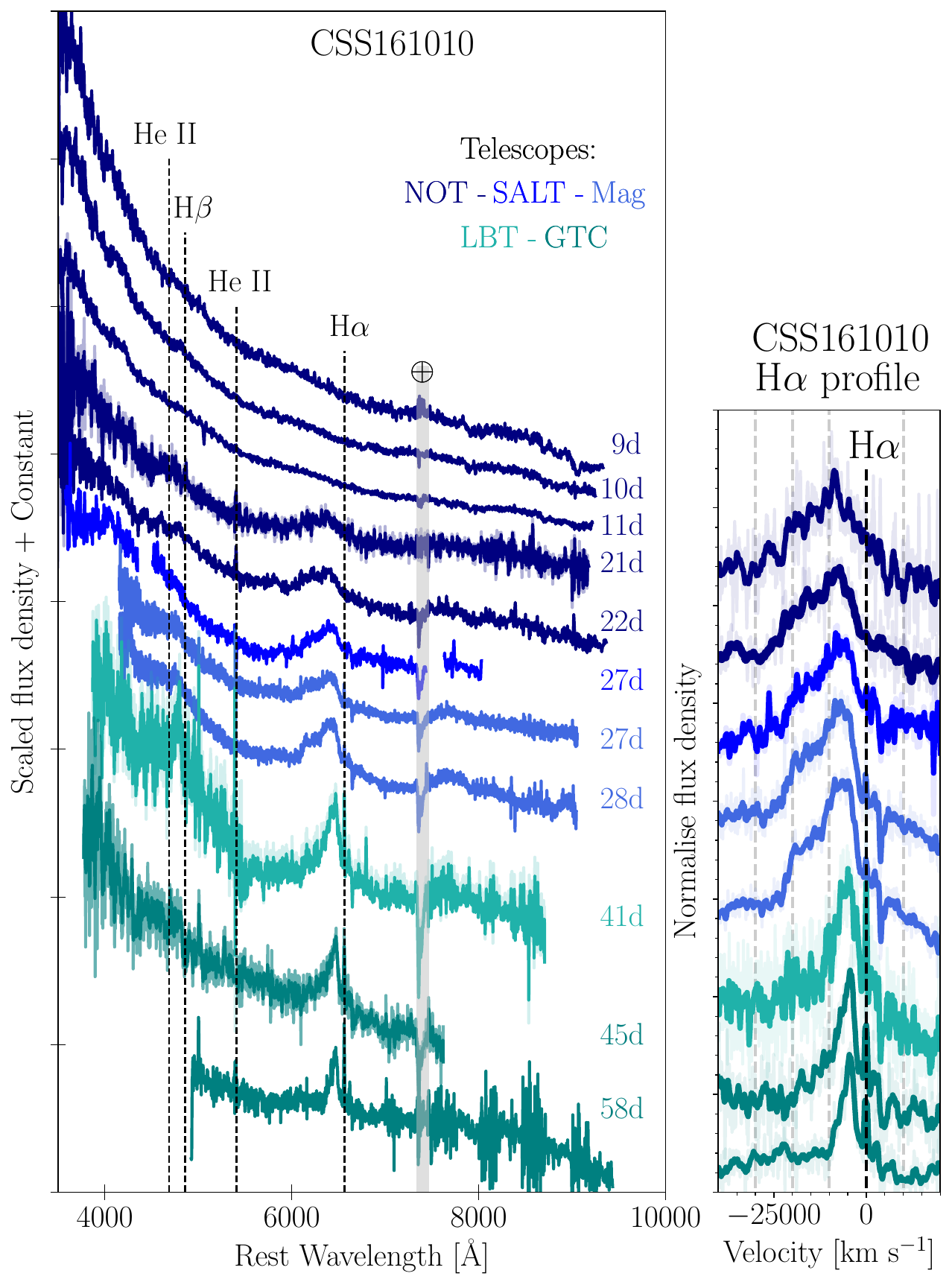}
\caption{Spectral sequence of \css\ from 9.4 to 57.7 days from the start of the outburst (JD = $2457669.92\pm2.00$). The phases are labelled on the right. Each spectrum has been corrected for MW extinction and shifted vertically for presentation. A zoom-in around the \ha\ P-Cygni profile in velocity space from 21 to 58 days is shown in the smaller right panel. Note the blueshift of \ha\ at all epochs.
}
\label{fig:css_spec}
\end{figure*}

Figure~\ref{fig:css_spec} presents the optical spectroscopic evolution of \css\ from 9.4 to 57.7 days from the start of the outburst. The spectra show a remarkable transformation from a blue and featureless continuum to spectra dominated by very broad, blueshifted emission lines with peculiar shapes and extremely high velocities. 

\begin{figure}
\centering
\includegraphics[width=\columnwidth]{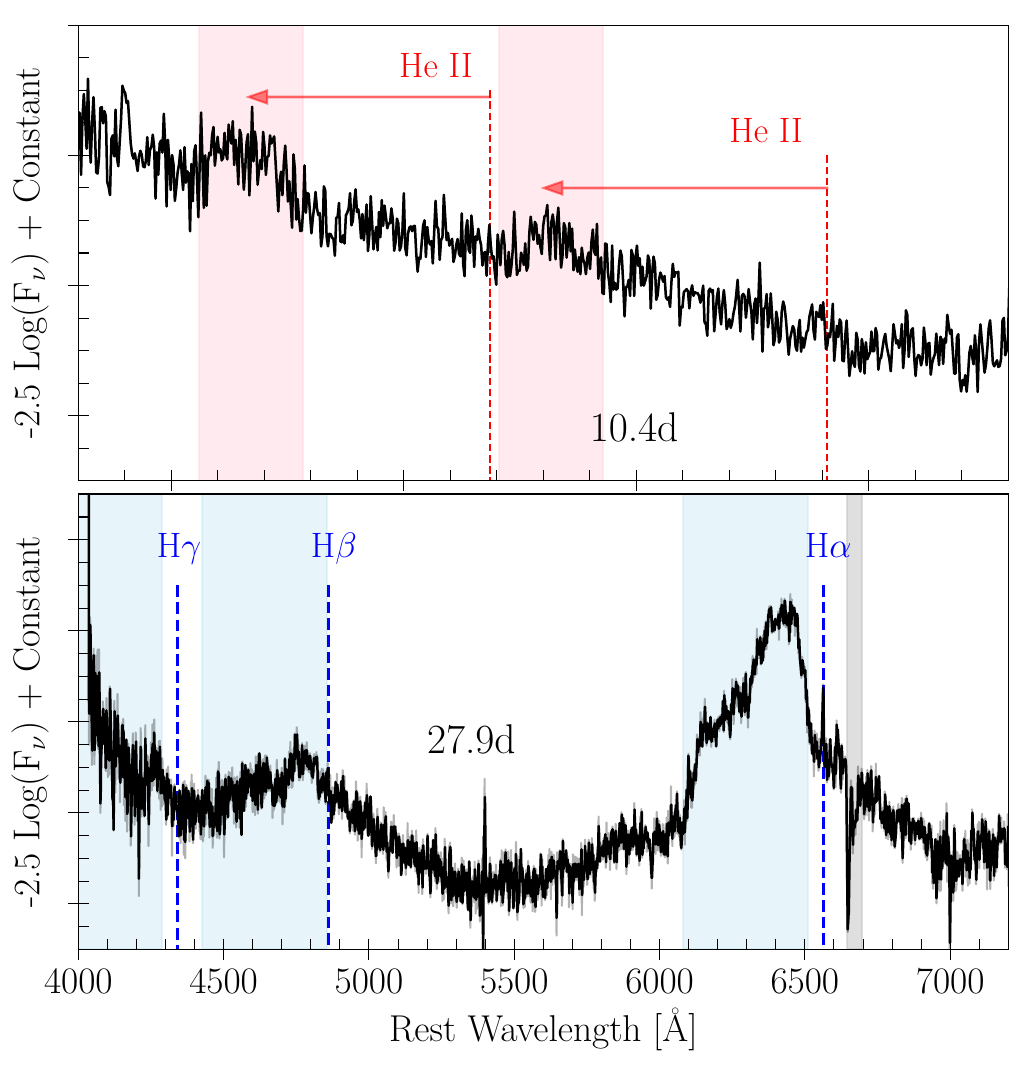}
\caption{Lines identified in the spectra of \css\ at $\sim10.4$ and $\sim27.9$ days post-start of the outburst. The dashed lines indicate the rest position of the helium (top) and hydrogen (bottom) lines. The shade regions mark their blueshifted locations.}
\label{fig:idlines}
\end{figure}

\begin{figure*}
\includegraphics[width=\textwidth]{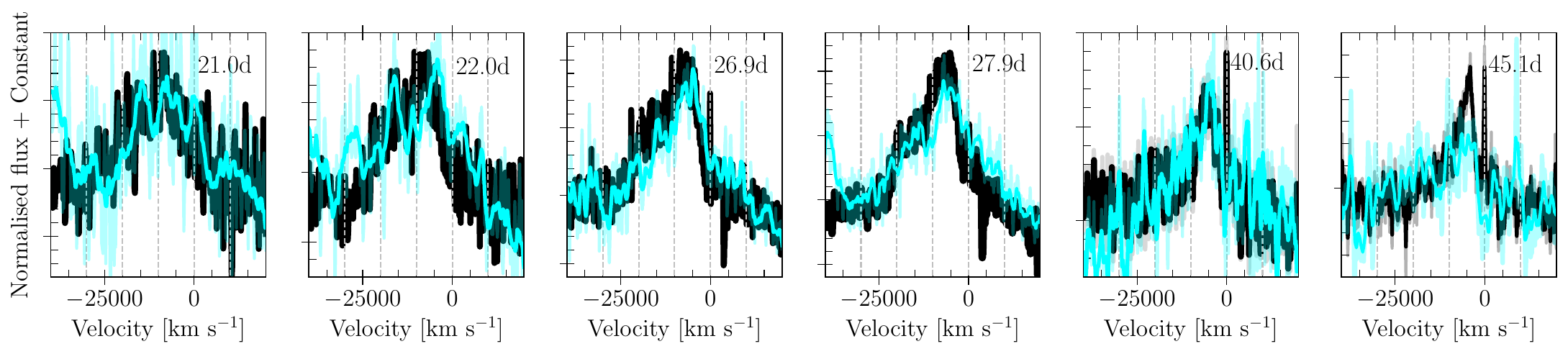}
\caption{Comparison of the H$\alpha$ (black) and H$\beta$ (cyan) profiles. from 21.0 to 57.7 days from the start of the outburst. The phases are labelled on the top of each panel. The vertical dashed lines mark the velocities at $-30000$, $-20000$, $-10000$, 0, and $10000$ \kms. To match both profiles, we multiplied H$\beta$ by 3 in the first four panels and by 2 in the last two. }
\label{fig:profiles}
\end{figure*}

The first spectrum, at 9.4 days, is characterised by a featureless blue continuum with a blackbody temperature of around T$_{BB}\simeq16000–17000$ K. One day later (10.4 days), the blackbody temperature decreased by $\sim500$ K, and the spectrum started to show some features between 4000 and 5000 \AA. After a detailed inspection, we found that these can be explained by \heii\ $\lambda4686$ and $\lambda5411$ emission lines at a velocity of $\sim-33000$ \kms\ (top panel of Figure~\ref{fig:idlines}). These lines are also visible at 11.3 days, but they are no longer detectable at 21.0 days. From 21.0 days onwards, the spectra changed completely, and two broad emission features at $\sim4500-5000$ \AA\ and $\sim6000-6500$ \AA\ are now visible. These lines show an extraordinary evolution over time: they lack any absorption component, and their emission components evolve from a broad feature into a more complex, flat-topped profile with a blue shoulder (at $\sim26-28$ days) and finally, after 40 days, into a feature with a narrower asymmetric shape. We identify these two broad emission features as \ha\ and \hb\ with blueshifted velocities between 10000 \kms\ at $\sim21$ days and 4000 \kms\ at $\sim58$ days (offset velocities from the rest wavelength at the broad line profile maxima). To verify this identification, we compare their profiles and evolution. Figure~\ref{fig:profiles} shows this comparison from 21 to 45 days. Although the lower signal-to-noise in the blue part of the spectra prevents a detailed analysis of \hb, we find that both profiles evolve consistently, confirming our initial identification of these lines. We found that the Balmer decrement is $\sim 3$ until 28 days, and perhaps a little smaller at later epochs.

The highest velocities of \ha\ and \hb\ (the bluest parts of the profile) decrease from $\sim-33000$ \kms\ at 21.0 days to $\sim-10000$ \kms\ at 57.7 days (Appendix~\ref{ap:ha}, Figure~\ref{fig:Hameasurements}), while the line centre always remains blueshifted by more than $4000$ \kms. The decreasing emission at the highest velocities is not unexpected since it reflects the disappearance of the fastest moving material; however, the persistent blueshifted emission and the lack of any emission at the rest wavelength (or redshifted from it) are striking. To our knowledge, this type of evolution has never been seen in any other transient.

\subsection{Comparison to other LFBOTs}
\label{sec:fbots}

\begin{figure*}
\centering
\includegraphics[width=\textwidth]{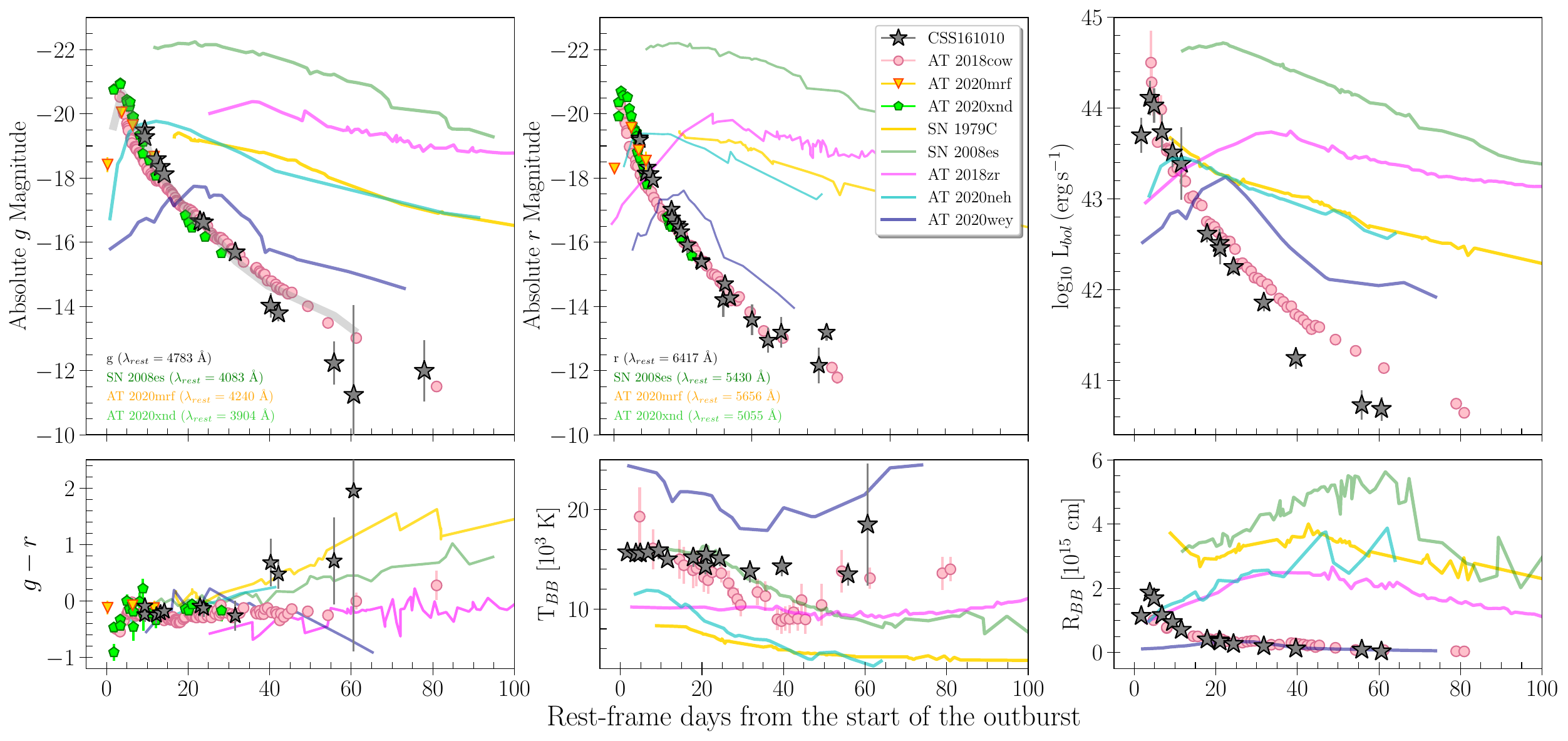}
\caption{
\textbf{Top panels:} \css\ $g-$band (left), $r-$band (middle) and bolometric light curves (right) compared with the well-sampled LFBOT AT~2018cow (purple circles), SNe with hydrogen in their spectra: SN~1979C (SN~II), SN~2008es (superluminous SN-II) and three hydrogen-rich TDE candidates: \zr, \neh\ and \wey. In the $g-$ and $r-$band light curve comparisons, the LFBOTs with good optical coverage AT~2020mrf (yellow down-facing triangles) and AT~2020xnd (green pentagons) are also included. The \css\ V-band light curve is also shown (grey thick line; left panel). Details of the comparison sample are presented in Table~\ref{tcomp}. \textbf{Bottom panels:} $g-r$ colours (left), blackbody temperature (middle) and blackbody radius (right) of \css\ compared with SN~1979C, SN~2008es, AT~2018cow, \zr, \neh\ and \wey. JD = $2457669.92\pm2.00$ is the start of the outburst estimated for \css.
}
\label{fig:compLC}
\end{figure*}

The fast photometric evolution of \css\ is reminiscent of other LFBOTs. Here, we compare our photometric and spectroscopic observations with those available for LFBOT objects. The best-observed cases are \cow, \mrf\ and \xnd. Information on these objects is presented in Table~\ref{tcomp}. In Figure~\ref{fig:compLC}, we show the $g-$ and $r$-band light curves, the bolometric light curve, $g-r$ colours, blackbody temperature and blackbody radius of \css\ together with those of LFBOTs. From the light curves, we can see that \css\ and \mrf\ have almost identical rise times of 3.8 and 3.7 days, respectively, whereas \cow\ has the fastest rise time of only 2.5 days. The rise time of \xnd\ is uncertain due to the poor constraints on its pre-peak light curve \citep{Perley21}. However, it seems to rise between 2 and 5 days. At peak, the brightest object is \xnd\ (M$_{5000}^{max}\simeq-20.9\pm0.3$ mag), followed by \cow\ (M$_{4800}^{max}\simeq-20.8\pm0.2$ mag), \css\ (M$_{5500}^{max}=-20.7\pm0.1$ mag) and \mrf\ (M$_{4200}^{max}\simeq-20.0\pm0.2$ mag)\footnote{\css\ and \mrf\ were only observed in one band around the peak.}. After the peak, they have similar decline rates and share similar blue $g-r$ colours up to $\sim30$ days post-outburst, when a deviation toward redder colours is observed in \css.

\begin{figure*}
\centering
\includegraphics[width=\textwidth]{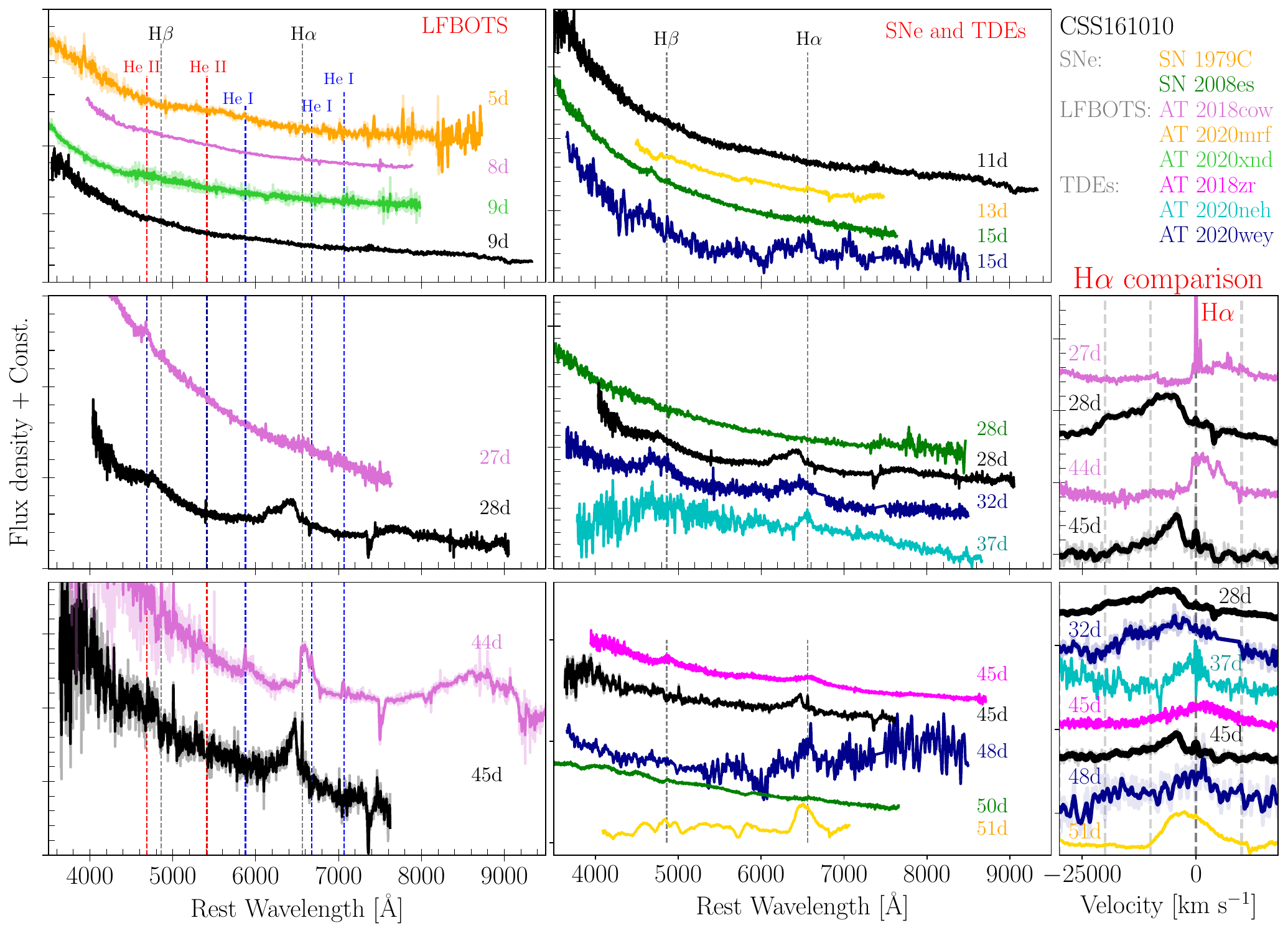}
\caption{\textbf{Left panels:} Spectral comparison of \css~ with the \cow-like objects: AT2018cow, AT2020xnd and AT2020mrf at different epochs. \textbf{Middle panels:} Spectral comparison of \css~ with fast-declining hydrogen-rich events: SNe~1979C and 2008es, and the TDE candidates \zr, \neh\ and \wey. Each spectrum has been shifted in flux for comparison. The vertical dashed lines indicate the rest wavelength positions of a selection of spectral lines. \textbf{Right panels:} \css\ \ha\ profile in velocity space compared with \cow\ (top) and with the fast-declining hydrogen-rich SN~1979C, and the TDE candidates \zr, \neh\ and \wey\ (bottom). Vertical dashed lines in the middle and right panels mark the velocities at $-20000$, $-10000$, 0, and $10000$ \kms.
}
\label{fig:spectralcomp}
\end{figure*}

Figure~\ref{fig:spectralcomp} shows the spectra of \css\ at three different phases compared with these LFBOTs (left panels). Before 10 days after the start of the outburst, all spectra are characterised by a featureless blue continuum. After this phase, where we only have spectroscopic data for \cow\ and \css, considerable differences appear. After $\sim20$ days, the spectrum of \css\ shows broad blueshifted hydrogen emission profiles, while \hei\ lines dominate the spectrum of \cow. Later, at $\sim45$ days, the emission lines of both \css\ and \cow\ become narrower, more significantly so in the former.

\subsection{Comparison to hydrogen-rich SNe and TDEs}
\label{sec:comp_sntde}

One of the main characteristics of \css\ is the presence of hydrogen lines in its spectra. Based on this property, we compare \css's light curves and spectra with hydrogen-rich SNe (SNe~II) and TDE candidates in Figures~\ref{fig:compLC} and \ref{fig:spectralcomp}. We selected well-studied objects of each of these classes; these are SN~1979C (fast-declining SN~II with a shallow \ha\ absorption feature; \citealt{Panagia80}), SN~2008es (SLSN-II;  \citealt{Gezari09}) and three hydrogen-rich TDE candidates: \zr\ \citep{Holoien19, Charalampopoulos22}; \neh\ \citep{Angus22} and \wey\ \citep{Charalampopoulos23}. Details of the comparison sample are presented in Table~\ref{tcomp}. From their photometric properties (Figure~\ref{fig:compLC}), we see that \css\ differs entirely from these hydrogen-rich events. The photometric evolution of hydrogen-rich SNe and TDE candidates is much slower, and their intrinsic colours are redder than those of \css. They also have lower ${\rm T}_{BB}$ (except for \wey; Figure~\ref{fig:compLC}) and larger ${\rm R}_{BB}$. Spectroscopically (Figure~\ref{fig:css_spec}), all objects are characterised by a featureless blue continuum in the early phases, but they begin to differ as the lines appear. Although SN~1979C and \css\ both show \ha, their profiles are dissimilar. The \ha\ profile of SN~1979C is bell-shaped, with a slightly blueshifted peak emission ($<-2000$ \kms). Blueshifted emission line peaks evolving to become rest frame-centred is a known property in SNe \citep{Dessart05, Anderson14a}. In the case of the SLSN-II SN~2008es, the \ha\ appears later (after $\sim100$ days \citealt{Gezari09}) and, during the comparison phases, shows different spectroscopic properties from \css. 

When comparing \css\ with TDE candidates, we see a large diversity. Unlike \css, \zr\ and \neh\ have an \ha\ profile centred at the rest wavelength; \zr\ has a symmetric line profile, but \neh\ does not. Both objects lack absorption features. For \neh, blueshifted lines were detected at late phases ($\sim212$ days from the start of the outburst) and linked to optically thick outflowing material \citep{Angus22}. Similar blueshifted profiles at very late times have been observed in SNe~II and attributed to dust formation (e.g. SN~1998S, SN~2007od; \citealt{Leonard00, Andrews10}). Regardless of this, their line profiles are different from \css. The spectral comparison with \wey\ is more interesting. At around 28-31 days, the spectra of \css\ and \wey\ appear similar. In particular, their \ha\ profiles are almost identical. The only difference is the missing redder part of the line profiles of \css\ (i.e. the entire profiles are blueshifted). However, later, at $45-48$ days, these two objects show very different spectra. TDEs often show blueshifted hydrogen line profiles with peculiar shapes \citep{Roth18}. This property is also observed in \css, and from all the objects included in the comparison sample, the hydrogen-rich TDE \wey\ is the most similar, although it has a distinctly slower luminosity evolution and a much larger $R_{BB}$ after $\sim 10$ days.

\subsection{Discussion} 
\label{sec:discus}

Some well-studied LFBOTs were found in dwarf star-forming galaxies \citep[e.g.][]{Coppejans20, Ho20, Yao22}, and thus have been considered likely to be associated with massive stars. We found that \css's host has very similar properties to \xnd's host, both have a small stellar mass (log(M$_{*}/$\Msun)$=8.13$ and 8.48) and a modest star formation rate (SFR(\Msun/yr)$=0.015$ and 0.020, respectively). Recently, two LFBOTs were found in more massive galaxies. \citet{Ho23} found that AT~2022tsd was located at $\sim6$ kpc from the centre of a star-forming galaxy (log(M$_{*}/$\Msun)$=9.96$), while \citet{Chrimes24, Chrimes24a} found that AT~2023fhn is 16.5 kpc from the centre of the nearest spiral galaxy and 5.4 kpc from an apparent dwarf companion (log(M$_{*}/$\Msun)$=9.97$ considering the spiral and satellite galaxies together), both objects representing a deviation in terms of their environments from previous LFBOTs. Both \citet{Ho23} and \citet{Chrimes24a} favoured a core-collapse event to explain these transients, although they did not rule out the IMBH TDE interpretation.

\subsubsection{Stellar explosion scenario}

\citetalias{Coppejans20} found that \css's X-ray and radio observations alone can be explained equally well by a stellar explosion or a TDE. Based on our optical observations, we find a stellar explosion to be unlikely. The presence of entirely blueshifted emission line profiles throughout the evolution is challenging to explain in any SN scenario. The ${\rm R}_{BB}$ evolution is also inconsistent with the homologous expansion expected in any SN \citep{Liu18}. Furthermore, the radioactively powered SN mechanism is unrealistic. Assuming that \css\ arises from a stellar explosion where all the energy is from radioactive decay, we can estimate the ejecta mass and the amount of \nifs\ synthesized during the explosion by fitting the Arnett model \citep{Arnett82}. Thus, considering a rise time (3.8 days), a canonical SN kinetic energy ($\sim10^{51}$ erg), an opacity of $\sim0.1$ cm$^{-2}$ g$^{-1}$ and  $\beta=13.7$ \citep[see,][]{Prentice16}, we derive an ejecta mass, M$_{ej}\sim0.3$ \Msun. Since \nifs\ powers the main peak, and using the rise time and the peak luminosity of the bolometric light curve, we estimate a \nifs\ mass of ${\rm M}_{\rm Ni}=2.2$ \Msun. The small M$_{ej}$ explains the fast rise, while the large amount of \nifs\ explains the peak luminosity. Given that ${\rm M}_{Ni}$ is much larger than the M$_{ej}$, this scenario would be unphysical. Additionally, considering that the spectra show prominent hydrogen lines, the M$_{ej}$ must also contain some hydrogen, making this inconsistency even greater.

Recent studies \citep[e.g.][]{Fox19, Xiang21, Pellegrino22} have found similarities between LFBOTs and Type Ibn SNe and suggest that these fast-evolving objects can be explained by the ejecta interacting with helium-rich circumstellar material (CSM). \citet{Xiang21} found that the bolometric light curve of \cow\ can be fitted by a hybrid model that includes \nifs\ and the interaction of the SN ejecta with a dense CSM. The spectra of \cow\ were argued to support this alternative scenario because they are dominated by narrow helium emission lines \citep{Fox19, Pellegrino22}. \cow\ and \css\ have similar light curves (Figure~\ref{fig:compLC}), so we might expect also \css\ to have a comparable powering mechanism (\nifs\ plus CSM). However, unlike \cow, the spectra of \css\ are dominated by very broad and entirely blueshifted hydrogen features. In \css, we only detect narrow lines at late phases, and those lines are associated with the host galaxy. Therefore, the spectral properties of \css\ do not support these scenarios.

\subsubsection{Wolf-Rayet/Black Hole Mergers}

The Wolf-Rayet star/black hole (or neutron star) merger scenario recently proposed for \cow\ \citep{Metzger22} seems problematic for \css\ because of the strong, blueshifted hydrogen lines. While the scenario allows the presence of hydrogen in a pre-SN disk-like structure, the orbital velocity of this disk is at least an order of magnitude slower than the observed $c/10$. Acceleration by an outflow from the central engine would be needed, combined with rapid cooling of the shocked gas to allow for efficient Balmer line emission. The situation could be similar to that of a cool, dense shell in SNe interacting with a dense surrounding media \citep[e.g.,][]{CheFra94}. The shocked disk would have a density of $\gtrsim 10^{9}~(T/10^8~{\rm K})^{0.5}$ cm$^{-3}$ ($T$ is the temperature of the shocked gas) to have a cooling time of $\lesssim 10$ days \citep[cf.][]{FLC96}. The gas cools under compression so that H$\alpha$-emitting gas would be orders of magnitude denser. Under these circumstances, Balmer decrements of $\sim 10$ are not unusual \citep[e.g.,][]{Taddia20} in contrast to the case B ratio of $\sim3$ we observe. It is also unclear why the merger scenario should result in strong blueshifts. 

The Case B \ha\ emission per cm$^{-3}$ is $\varepsilon_{{\rm H}\alpha} = 3.51\times10^{-25} T_4^{-0.96} n_e n_p f^2$ \citep{OF06}, and if $V_{\rm em}$ is the volume of the gas emitting \ha, then the \ha\ luminosity is $L_{{\rm H}\alpha} = \varepsilon_{{\rm H}\alpha} V_{\rm em}$. Here $T_4$ is the gas temperature in $10^4$ K, $f$ is the filling factor of the emitting gas, and $n_e$ and $n_p$ are the electron and proton number densities, respectively. For a helium-to-hydrogen number density ratio of 0.1 and fully ionized hydrogen, the mass of this volume is $M_{\rm em} = 1.4 m_p n_p f V_{\rm em}$, or 
$M_{\rm em} = 0.33~T_4^{0.96}~(f n_{e,8})^{-1}~L_{{\rm H}\alpha,40}$~\Msun, where $n_{e,8}$ is in units of $10^8$ cm$^{-3}$ and $L_{{\rm H}\alpha,40}$ in $10^{40}$ erg s$^{-1}$. For an ejected mass of 0.1~\Msun~and a temperature of $10^4$ K, this means that $n_{e,8}\gtrsim 3.3 L_{{\rm H}\alpha,40}$. We measured $L_{{\rm H}\alpha,40} \approx 0.3$ when \ha\ was first detected on day 21, and peaking at $L_{{\rm H}\alpha,40} \approx 1.7$ one week later, translating into $n_{e,8}\gtrsim 0.95$, and $\gtrsim 5.6$, respectively. 

The Case B recombination time of hydrogen is $(\alpha_{\rm B} n_e)^{-1}$, where $\alpha_{\rm B}= 2.59\times10^{-13} T_4^{-0.86}$~cm$^{-3}$~s$^{-1}$ \citep{OF06}, is less than one day for $n_{e,8}\gtrsim 0.45 T_4^{0.86}$. This implies a steady state between ionisation and recombination, potentially with a time lag. If $\dot N_{{\rm ion},50}$ is the number of ionising photons per second in units of $10^{50}$ s$^{-1}$ put out by the central source, and $\Omega/4\pi$ is the solid angle subtended by the H$\alpha$-emitting region as seen by the central source, $\dot  N_{{\rm ion},50} \approx 74 (\Omega/4\pi)^{-1} T_4^{0.1} L_{{\rm H}\alpha,40}$. Here we have assumed that the total Case B recombination rate in the \ha-emitting region is $\alpha_{\rm B} n_e n_p f^2 V_{\rm em}$. If we use $L_{{\rm H}\alpha,40}$ from 21 and 28 days, then the corresponding values for $ (\Omega/4\pi) \dot  N_{{\rm ion},50}$ are $\approx 21$ and $\approx 120$, respectively. This is in stark contrast to what is expected from black-body spectra with the properties in Figure~\ref{fig:compLC}. At 21~(28) days black bodies only generate $\dot  N_{{\rm ion},50} \sim 15~(3.6)$. A possible solution is that there is a $\sim 3$ week delay between the emitted ionising radiation and \ha\ emission due to geometry and light travel time, so that the number of ionising photons reaching the H$\alpha$ cloud on day 28 may correspond to an epoch close to the peak of the bolometric luminosity in Figure~\ref{fig:compLC}. The luminosity is then some $\sim 35$ times higher than on day 28, and $\dot  N_{\rm ion} \sim40$ times larger (for a black-body spectrum). A 3-week delay between continuum and \ha\ emission geometrically would mean that the distance between the \ha-emitting regions and the central source is $\gtrsim 8\times10^{16}$~cm. 
Similar arguments for light-travel time effects in TDEs are discussed by \cite{Charalampopoulos22}. A problem with this interpretation is that $(\Omega/4\pi)$ is expected to be much less than unity for such a distance, and that we see \heii\ lines much earlier than H$\alpha$. 

It is, therefore, likely that a pure black-body spectrum from a central source cannot produce enough ionising radiation. One obvious candidate for the 'extra' ionising radiation is the X-ray emission observed by \citetalias{Coppejans20}, which for their assumed spectrum with energy distribution $\propto \nu^{-1}$ had a luminosity of $\sim 4\times10^{39}$ erg s$^{-1}$ between 0.3-10 keV $3-4$ months after optical peak. Extrapolating to lower energies, this would mean a rate of the number of ionising photons from the X-ray source that is $N_{{\rm ion},X,50} \sim 0.5$. If the process is similar to that in active galactic nuclei (AGN; i.e., inverse Compton scattering), the number of ionising photons from the X-ray source is roughly proportional to the optical/UV luminosity. This could mean $N_{{\rm ion},X,50} \gtrsim 100$ close to the optical peak, which would still only account for $\sim 10^{-3}$ of the total number of black-body photons. Interestingly, the X-ray properties for \cow\ derived by \cite{Margutti19} indicate that it had $N_{{\rm ion},X,50} \sim 300$ around 10 days. The upper limit from the stacked Swift XRT data for \css\ at 29.5 days (cf. Appendix~\ref{ap:photo}) corresponds to $N_{{\rm ion},X,50} \leq 31$, assuming a flux distribution that is $\propto \nu^{-1}$. With the evolution of \cow\ as a template, this would mean $N_{{\rm ion},X,50} \lesssim 70~(200)$ at 10 days (and at peak) for \css. There is thus room for sufficient X-ray emission to ionise the line-emitting gas.

A picture emerges where the fastest line-emitting gas expands with a speed of $\sim c/10$, and that it is, like in AGN, ionised mainly by the X-ray emission. After 11~(28) days when weak \heii\ lines are first seen (and when \ha\ is the strongest) this gas has reached $\sim 3~(7)\times10^{15}$ cm. Light travel time can only account for a few days of delay if the ionising source is central. We find it more likely that the X-rays are produced over a larger volume, possibly in a jet-like structure as expected in TDEs \citep[see, e.g.,][]{Charalampopoulos22}. The extent of the X-ray emitting region can be estimated from the expansion of the mildly relativistic blast wave in the models of \citetalias{Coppejans20} for the same epochs, which are $\sim 2~(4.5)\times10^{16}$ cm. This picture seems incompatible with the Wolf-Rayet star/compact object merger scenario. However, it is consistent with a TDE where time-lags between the broadband light curves and the fluxes of broad emission lines have been observed, with shorter lags for, e.g., \hei\ than \ha\ \citep[see, e.g.][]{Charalampopoulos22, Faris24}.

\subsubsection{TDE from an IMBH scenario}
\label{tdeimbh}

For a number of reasons, we find that a TDE appears to be the least problematic explanation for the spectroscopic and photometric properties of \css, and also fits with the arguments about \ha\ and light-travel times above. 

The entirely blueshifted line profiles of \css\ at all epochs can be explained as a result of outflows occurring in TDEs as previously seen in X-rays \citep{Kara16} but also in blueshifted UV/optical emission/absorption lines \citep{Hung19, Nicholl20}. In fact, \cite{Coppejans20} concluded that \css\ has a mildly relativistic, decelerating outﬂow with an initial velocity of $\Gamma\beta c > 0.55c$, that decreases to $\sim c/3$ after one year. Some TDEs are known to launch a relativistic jet, e.g., the well-observed case Swift J164449.3+573451 \citep{Burrows11}. In the case of Arp~299-B~AT1, interpreted as a TDE, VLBI observations revealed an expanding and decelerating radio jet with an average intrinsic speed of 0.22c \citep{Mattila18}, which is quite similar to the outflow speed estimated for \css\ at the late times. 

Furthermore, the peak bolometric luminosity of \css, L$_{bol}=1.30~(\pm 0.56) \times 10^{44}$ \ergs, implies a highly super-Eddington accretion phase (for a BH mass of $10^{2.93}$ -- $10^{4.77}$ \Msun, the Eddington luminosity is L$_{Edd}= (1-70) \times 10^{41}$ \ergs). Although very highly super-Eddington accretion can be expected in TDEs \citep{Skadowski16}, the effect is not expected to be as large for the optical luminosities, making the lowest BH masses less likely. In any case, strong outflows are expected \citep{Shakura73, Lipunova99} and for a TDE produced by an IMBH, we expect these outflows to be much stronger and persist longer than for a supermassive BH because the super-Eddington mass fallback to the BH is predicted to persist for extended periods \citep{Wu18}. Moreover, the lack of any absorption component in the optical lines can also be interpreted as a result of an outflow \citep{Roth18}. The disappearance of the bluest part (highest velocity) of the \ha\ profile may also be explained as the result of an outflow that expands with time \citep{Roth18}. 

The continuous decay of ${\rm R}_{BB}$ after only $\sim3-4$ days from the start of the outburst can be described within the framework of TDEs \citep{Liu18}. Here, the ${\rm R}_{BB}$ of \css\ is within the typical values found for TDEs (\citealt{Holoien19, Charalampopoulos22, Charalampopoulos23}; $\sim10^{14}$ – $10^{15}$ cm). Moreover, the fast light curve decline rates could be explained as a partial TDE, in which a stellar remnant survives the interaction. Simulations have shown that the mass fallback rate to the BH in a partial TDE differs substantially from a t$^{-5/3}$ power law expected for a total tidal disruption \citep{Rees84}, and decays with a steep power law of anywhere from t$^{-2}$ to t$^{-5}$ \citep{Coughlin19, Ryu20}. Hydrodynamic simulations \citep{Kirouglu23} of a close encounter between a 1 \Msun\ main-sequence star and a $10^{2} - 10^{4}$ \Msun\ IMBH found that a small amount of material is stripped from the star at each pericentre passage with a period of a few $\times 10^{3}$ yr for a $10^{4}$ \Msun\ BH. The star is eventually fully disrupted or ejected. The simulations predict highly super-Eddington accretion rates and brief flares with peak luminosities between $10^{44}$ erg s$^{-1}$ and $3 \times 10^{44}$ \ergs\ for BH masses in the range $10^{3}-10^{4}$ \Msun. These predicted peak luminosities are similar to that observed for \css, providing additional support for this scenario.

\begin{figure}
\centering
\includegraphics[width=\columnwidth]{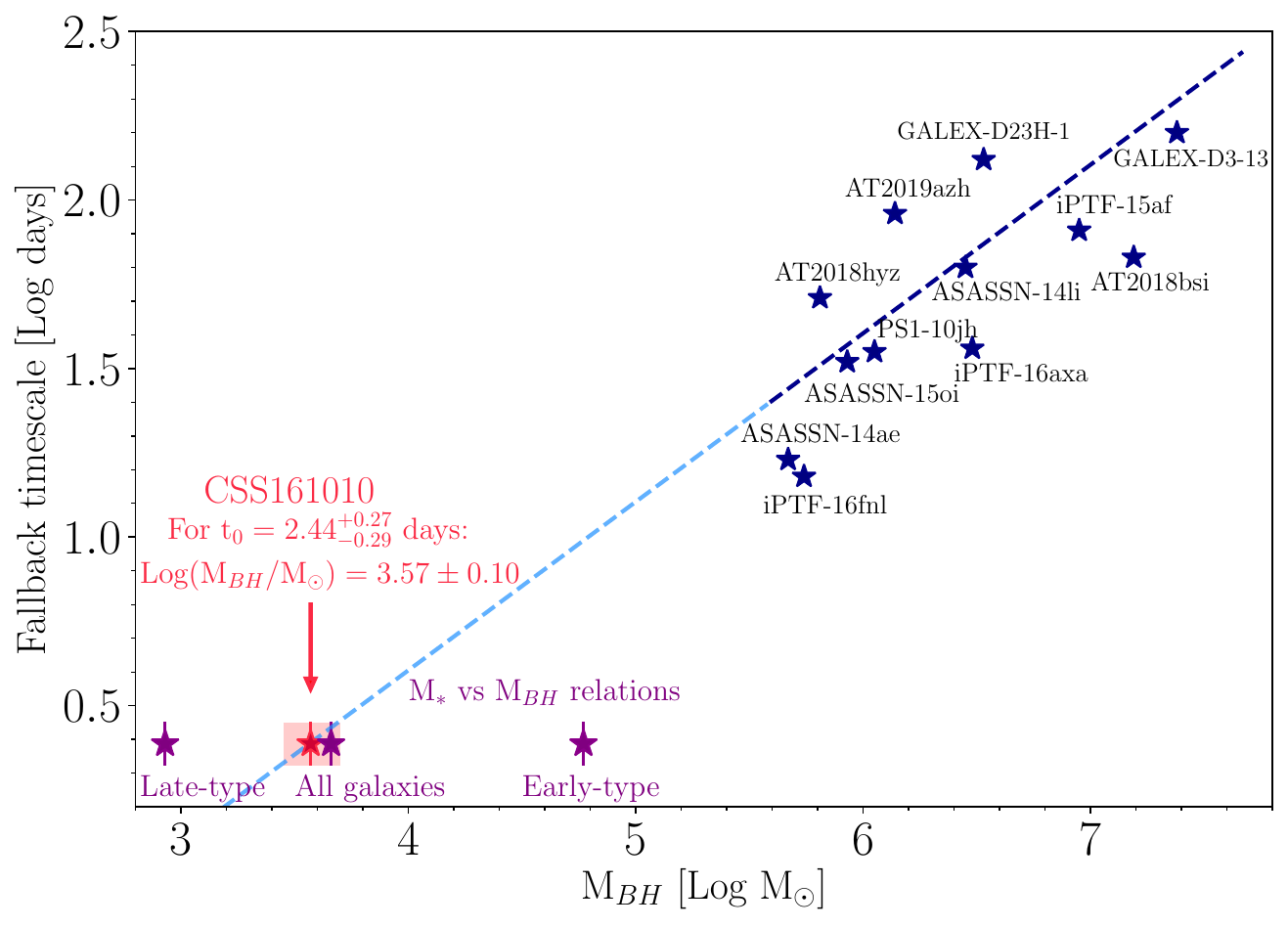}
\caption{
Correlation between the black-hole mass and TDE light curve decay time \citep{vanVelzen20}. The dashed blue line shows the expected relation between ${\rm t}_{\text{fb}}$ and ${\rm M}_{\text{BH}}$ for a star of one solar mass. This line is extended (light blue) to lower values. 
The red star marks the location of \css\ obtained by extending the ${\rm t}_{\text{fb}}$ and ${\rm M}_{\text{BH}}$ correlation. The red rectangle shows the uncertainty obtained from the light curve fitting (Appendix~\ref{ap:tde}, Figure~\ref{fig:LCfits}). Purple stars show the BH masses obtained from the ${\rm M}_{*}$ and ${\rm M}_{\text{BH}}$ relations (Figure~\ref{fig:hostgal}; Appendix~\ref{ap:host}). Reproduced and modified from \citep{vanVelzen20}.
}
\label{fig:BHM_FBD}
\end{figure}

The fallback timescale (t$_{\text{fb}}$) measured from the TDE's light curve can provide an independent BH mass estimate \citep{Blagorodnova17, vanVelzen20, Mummery23}. This fallback timescale has been estimated for a sample of TDEs by fitting a power-law decay (L$_{\text{bb}}\propto\text{(t/t}_{0})^{p}$) to their bolometric light curves with the power-law index fixed, $p=-5/3$. As the characteristic decay time (t$_0$) is comparable to the theoretical t$_{\text{fb}}$ \citep{vanVelzen19}, it is possible to derive the BH mass. Fitting the bolometric light curve of \css, we found t$_0=2.44^{+0.27}_{-0.29}$ days (Appendix~\ref{fig:LCfits}), which implies a BH mass of $10^{3.57\pm0.10}$ \Msun\ (Figure~\ref{fig:BHM_FBD}) consistent with our estimates based on the host galaxy's stellar mass; this corroborates \css\ as a TDE candidate with one of the lowest BH masses to date. 

Most low-redshift IMBH candidates have been found in low-mass, star-forming dwarf galaxies \citep{Mezcua16} through kinematic studies or by extrapolating the scaling relations between the BH properties and galaxy parameters. Observations of AGN, gravitational-wave signals, and TDEs have provided further evidence for their existence \citep{Donato14, Mezcua17, Lin18, Greene20, Abbott20, He21, Wen21, Angus22}. Although an IMBH is not expected to exist in every dwarf galaxy, simulations \citep{Bellovary19} indicate that the BH occupation fraction rapidly increases in galaxies with stellar masses above $10^{8}$ \Msun, with roughly 60\% of $10^{8.3}$ \Msun\ galaxies hosting an IMBH. For \css, we measured an offset of $0\farcs383 \pm 0\farcs024$ (a projected distance of $\sim$ 300 pc; Appendix~\ref{ap:location}) from the host galaxy's centre. Although super-massive BHs are located at the centres of their host galaxies, IMBHs in dwarf galaxies are not all expected to coincide with the galaxy nucleus \citep{Lin18}. In fact, recent simulations \citep{Bellovary19} indicate that half of the IMBHs in dwarf galaxies {are more than 400 pc from the centres}. BH growth in low-mass galaxies is expected to be stunted by SN feedback \citep{Habouzit17}, so dwarf galaxies in the local Universe that have evolved in isolation can host BHs with masses comparable to the seed BHs in the early Universe.

\section{Summary}
\label{sec:sum}

We presented photometric and spectroscopic observations of \css. We found that the light curves of \css\ are characterized by an extremely fast evolution and blue colours. \css\ reaches an absolute peak of M$_{V}^{max}=-20.66\pm0.06$ mag in 3.8 days from the start of the outburst. After-maximum, \css\ follows a power-law decline $\propto t^{-2.8\pm0.1}$ at all optical bands. These photometric properties are comparable to those shown by well-observed LFBOTs. However, unlike these objects, the spectra of \css\ are dominated by very broad blueshifted hydrogen emission lines starting at $\sim20$ days from the start of the outburst. Our analysis shows any stellar explosion scenario to be unlikely and that \css\ is most naturally explained as a hydrogen-rich star (partially) disrupted by an IMBH. Multiwavelength observations of other LFBOTs \citep{Pasham21, Zhang22, Inkenhaag23, Chen23, Ho23} provide strong evidence favouring a central engine. Although the nature of the engine is unknown, a TDE by an IMBH is a plausible scenario in several cases \citep{Inkenhaag23, Ho23}. In fact, we argue that from a spectroscopic point of view, \css\ provides the most convincing case to date. If other LFBOTs, in addition to \css\ and \cow, can be explained as a TDE by an IMBH, further observations of such events could be used for pinpointing otherwise quiescent IMBHs, constraining their masses, occupation fractions, host galaxy properties and galactocentric distances.


%\begin{acknowledgments}
\section*{acknowledgments}
We thank the anonymous referee for the comments and suggestions
that helped us improve the paper. 
We thank Keiichi Maeda, Masaomi Tanaka, Hanindyo Kuncarayakti, Takashi Nagao and Taeho Ryu for valuable discussions; Anna Ho and Daniel Perley for discussing the physical properties of LFBOTs during the \textit{Interacting Supernovae} MIAPbP meeting; Sjoert Van Velzen for sharing the data to reproduce Figure~\ref{fig:BHM_FBD}; Francesco Coti Zelati for helpful discussion about the X-ray observations and upper limits. We are grateful to Nidia Morrell for performing some of the observations used in this work.

C.P.G., S.M., L.D. thanks the Munich Institute for Astro-, Particle and BioPhysics (MIAPbP), which is funded by the Deutsche Forschungsgemeinschaft (DFG, German Research Foundation) under Germany´s Excellence Strategy – EXC-2094 – 390783311, for the useful discussions.   

C.P.G. acknowledges financial support from the Secretary of Universities and Research (Government of Catalonia) and by the Horizon 2020 Research and Innovation Programme of the European Union under the Marie Sk\l{}odowska-Curie and the Beatriu de Pin\'os 2021 BP 00168 programme, from the Spanish Ministerio de Ciencia e Innovaci\'on (MCIN) and the Agencia Estatal de Investigaci\'on (AEI) 10.13039/501100011033 under the PID2020-115253GA-I00 HOSTFLOWS project, and the program Unidad de Excelencia Mar\'ia de Maeztu CEX2020-001058-M. 
S.M. acknowledges financial support from the Research Council of Finland project 350458. P.L. acknowledges support from the Swedish Research Council. 
S.G.G. acknowledges support by FCT under Project~No.~UIDB/00099/2020.
P.G.J. has received funding from the European Research Council (ERC) under the European Union’s Horizon 2020 research and innovation programme (Grant agreement No.~101095973).
S.D. acknowledges the National Natural Science Foundation of China (Grant No. 12133005) and the New Cornerstone Science Foundation through the XPLORER PRIZE.
N.E.R. acknowledges partial support from MIUR, PRIN 2017 (grant 20179ZF5KS). N.E.R, A.P., S.B. acknowledge the PRIN-INAF 2022 grant "Shedding light on the nature of gap transients: from the observations to the models".
C.S.K. is supported by NSF grants AST-1908570 and AST-2307385 and AST-2407206.
M.F. is supported by a Royal Society - Science Foundation Ireland University Research Fellowship.
R.K. acknowledges support via the Research Council of Finland (grant 340613).
M.D.S. is funded by the Independent Research Fund Denmark (IRFD, grant number  10.46540/2032-00022B). 
J.L.P. acknowledges support from ANID, Millennium Science Initiative, AIM23-0001.
L.W. and M.G. acknowledge funding from the European Union's Horizon 2020 research and innovation programme under grant agreement No. 101004719 (OPTICON-RadioNET Pilot, ORP). L.W., M.G. and A.H. acknowledge support from the Polish National Science Centre (NCN) grant No 2015/17/B/ST9/03167 (OPUS) to L.W., and the Polish participation in SALT is funded by grant No. MNiSW DIR/WK/2016/07.
J.M. has the support of the National Key R\&D Program of China (2023YFE0101200), the National Natural Science Foundation of China (NSFC 12393813), CSST grant CMS-CSST-2021-A06, and the Yunnan Revitalization Talent Support Program (YunLing Scholar Project). 
E.R.C. acknowledges support from the National Research Foundation of South Africa.

Based on observations made with the Nordic Optical Telescope, owned in collaboration by the University of Turku and Aarhus University, and operated jointly by Aarhus University, the University of Turku and the University of Oslo, representing Denmark, Finland and Norway, the University of Iceland and Stockholm University at the Observatorio del Roque de los Muchachos, La Palma, Spain, of the Instituto de Astrofisica de Canarias.

Observations from the NOT were obtained through the NUTS collaboration, which are supported in part by the Instrument Centre for Danish Astrophysics (IDA). The data presented here were obtained in part with ALFOSC, which is provided by the Instituto de Astrofisica de Andalucia (IAA) under a joint agreement with the University of Copenhagen and NOTSA.

Based on observations made with the GTC telescope in the Spanish Observatorio del Roque de los Muchachos of the Instituto de Astrofísica de Canarias.

This work is based on observations made with the Large Binocular Telescope. The LBT is an international collaboration among institutions in the United States, Italy and
Germany. LBT Corporation partners are: The University of Arizona on behalf of the Arizona Board of Regents; Istituto Nazionale di Astrofisica, Italy; LBT Beteiligungsgesellschaft,
Germany, representing the Max-Planck Society, The Leibniz Institute for Astrophysics Potsdam, and Heidelberg University; The Ohio State University, representing OSU, University of Notre Dame, University of Minnesota and University of Virginia.

Some of the observations reported in this paper were obtained with the Southern African Large Telescope (SALT), as part of the Large Science Programme on transients 2016-2-LSP-001 (PI: Buckley).

This work has made use of data from the Asteroid Terrestrial-impact Last Alert System (ATLAS) project. The Asteroid Terrestrial-impact Last Alert System (ATLAS) project is primarily funded to search for near-earth asteroids through NASA grants NN12AR55G, 80NSSC18K0284, and 80NSSC18K1575; byproducts of the NEO search include images and catalogs from the survey area. This work was partially funded by Kepler/K2 grant J1944/80NSSC19K0112 and HST GO-15889, and STFC grants ST/T000198/1 and ST/S006109/1. The ATLAS science products have been made possible through the contributions of the University of Hawaii Institute for Astronomy, the Queen’s University Belfast, the Space Telescope Science Institute, the South African Astronomical Observatory, and The Millennium Institute of Astrophysics (MAS), Chile.

The Liverpool Telescope is operated on the island of La Palma by Liverpool John Moores University in the Spanish Observatorio del Roque de los Muchachos of the Instituto de Astrofisica de Canarias with financial support from the UK Science and Technology Facilities Council. This work makes use of data from the Las Cumbres Observatory network.

This research uses data obtained through the Telescope Access Program (TAP), which has been funded by the TAP member institutes.  We thank the former Swift PI, late Neil Gehrels, the Observation Duty Scientists, and the science planners for approving and executing our Swift/UVOT program.

This publication makes use of data products from the Wide-field Infrared Survey Explorer, which is a joint project of the University of California, Los Angeles, and the Jet Propulsion Laboratory/California Institute of Technology, funded by the National Aeronautics and Space Administration. This publication also makes use of data products from NEOWISE, which is a project of the Jet Propulsion Laboratory/California Institute of Technology, funded by the Planetary Science Division of the National Aeronautics and Space Administration.

We thank Las Cumbres Observatory and its staff for their continued support of ASAS-SN. ASAS-SN is funded in part by the Gordon and Betty Moore Foundation through grants GBMF5490 and GBMF10501 to the Ohio State University, and also funded in part by the Alfred P. Sloan Foundation grant G-2021-14192.

%\end{acknowledgments}


%\facilities{}
%\software{} 



\clearpage
\appendix

\section{Observations of \css}
\label{ap:obs}

\subsection{Photometry}
\label{ap:photo}

Multi-wavelength photometric coverage of \css\ was acquired between 2016 October 10 and 2016 December 27. During the first nine days, three epochs of $V$ photometry were obtained by ASAS-SN; after this, $BgVRrIiz$ optical imaging data were obtained with the 2-m Liverpool Telescope (LT) using the IO:O imager, the 2.56-m Nordic Optical Telescope (NOT) using the Alhambra Faint Object Spectrograph and Camera (ALFOSC) at the Roque de Los Muchachos Observatory (Spain), the 1.0-m telescopes of Las Cumbres Observatory Global Telescope Network (LCOGT; \citealt{Brown13}), the 2.4m Hiltner telescope at the Michigan-Dartmouth-MIT (MDM) Observatory, the 1.04m Sampurnanad Telescope in the Aryabhatta Research Institute of observational sciencES (ARIES), the 1.3-m Devasthal Fast Optical (DFOT; \citealt{Sagar12}) telescope at Nainital (India) and the imaging modes of the Low Dispersion Survey Spectrograph (LDSS-3) and Inamori Magellan Areal Camera and Spectrograph (IMACS) mounted on the 6.5-m Magellan telescopes. Additionally, seven epochs of near-infrared (NIR) $H$ photometry were obtained with the LT using the IO:I imager, while three epochs of UltraViolet (UV) Optical observations were obtained with the UltraViolet/Optical Telescope (UVOT) on board the Swift spacecraft. There were no X-ray detections associated with these Swift observations, with a 3-sigma Swift XRT upper limit on the 0.5-10 keV flux of $4\times10^{-13}$ erg cm$^{-2}$ at 25.14 days and $8\times10^{-14}$ erg cm$^{-2}$ at 29.5 days (stacking the data from 25.14, 30.17 and 33.06 days). Here we have assumed a flux distribution that is $\propto \nu^{-1}$, and a column density of X-ray absorbing gas consistent with only Milky Way absorption. All NOT observations were obtained through the NOT Unbiased Transient Survey (NUTS\footnote{\url{https://nuts.sn.ie}}) allocated time. 

All images were reduced using standard procedures, including bias removal and flat field correction. For the NIR images, the reductions also included sky subtraction. We used the photometric pipeline \texttt{PmPyeasy} \citep{Chen22} to obtain the optical and NIR photometry. We followed the photometry procedures outlined in \cite{Chen22}, which primarily include the following three steps: image registration and source detection, measuring instrumental magnitudes with arbitrary zero points, and deriving photometric zero points to put the magnitudes into standard magnitude systems.  We performed aperture photometry for all the images using a 5\farcs0 radius circular aperture.  We used a relatively large aperture to include the flux of both the transient and the underlying host galaxy of CSS161010. We used the Pan-STARRS1 DR1 MeanObject database \citep{Flewelling20} for the optical-band photometric calibration. The final $B, V$ magnitudes are in the Vega system, and $g, r, i, z$-band magnitudes are in AB magnitudes.  We used the 2MASS \citep{Skrutskie06} photometric catalogue for the NIR photometric calibration. The NIR magnitudes are in the Vega system. The host galaxy flux was subtracted to get the transient brightness. We used galaxy images obtained with the 6.5-m Magellan telescopes (LDSS-3 and IMACS) on 20170721 and 20170919 to estimate the host galaxy flux.  

We used the \texttt{HEAsoft}\footnote{\url{https://heasarc.gsfc.nasa.gov/docs/software/heasoft/} v. 6.29c.} toolset for Swift UVOT photometry.  We first summed the exposures for each epoch using the task \texttt{uvotimsum}, and then we extracted source counts from a 5\farcs0 radius region centred on CSS161010 using task \texttt{uvotsource}.  The source counts were converted into the AB magnitude system based on the most recent UVOT calibrations \citep{Breeveld11}. The galaxy flux was subtracted to get the Swift UVOT photometry. Optical, UVOT and NIR photometry are presented in Tables~\ref{tphoto}, \ref{tphotodfot}, \ref{tphotoswift} and \ref{tphotoltH}. 

Photometry in the orange ($c$) and cyan ($o$) filters (blue and red filters that cover a wavelength range between 4200 and 6500 \AA\ and from 5600 to 8200 \AA, respectively) was obtained by the twin 0.5 m Asteroid Terrestrial-impact Last Alert System (ATLAS \citealt{Tonry18, Smith20}) and through the ATLAS forced photometry server\footnote{\url{https://fallingstar-data.com/forcedphot/}}. Tabe~\ref{tphotoatlas} lists the mean magnitudes. 

The host galaxy of \css\ was also imaged by the Wide-field Infrared Survey Explorer (WISE) in multiple epochs before and after its discovery.

We also obtained two epochs of NIR imaging with NOTCam at the NOT on 2017 February 6 (in $H$) and 2017 February 19 (in $JK$). These images were reduced using standard NIR reduction techniques in the NOTCam package\footnote{https://www.not.iac.es/instruments/notcam/guide/observe.html} in \textsc{iraf}. We searched for IR emission from the transient using the NIR images obtained on 2018 November 15 \citepalias{Coppejans20}. After alignment and subtraction \citep{Alard98, Alard00} between the 2017 and 2018 images, we found no trace of emission from the transient. A bright NIR source near the position of \css\ was detected in the NOTCam images, which is the same source as reported in 2020 \citepalias{Coppejans20}. It does not show any significant variability between the two observations. \\

The bright NIR source is also detected in the Wide-field Infrared Survey Explorer (WISE) satellite in the W1 and W2 bands at 3.6$\mu m$ and 4.4$\mu m$, respectively \citepalias{Coppejans20}. We queried the unTimely catalogue \citep{Meisner23}, a time-domain catalogue of WISE detections derived from the unWISE co-added images \citep{Meisner18}, in order to search for any variability in the source that could indicate transient flux associated with CSS161010. The mean date of the images obtained by the WISE satellite in the first visit after the first ASAS-SN detection of CSS161010 was on 2017 February 17 (JD=2457801.93; phase = 130d). The mean W1 Vega magnitude of all the 16 individual detections from 2010 to 2020 in the unTimely catalogue is 16.73, with a dispersion of $\sigma=$0.09 mag. The W1 magnitude of the source in the 2017 February visit is 16.6$\pm$0.1 mag, consistent with no variability. The W2 detection of the source is marginal, and it is not detected in most of the individual unWISE co-adds, notably in the 2017 February visit. Finally, we performed image subtraction between the unWISE W1 images immediately before and after the transient using the same methods as above and found no residual that would indicate transient flux.

\subsection{Imaging polarimetry}

Observations of \css\ were made on 2016 December 2, during the commissioning of the imaging polarimetric mode of the Robert Stobie Spectrograph (RSS; \citealt{Burgh03}) at the South African Large Telescope (SALT; \citealt{Buckley06}). Observations were made using the PI06645 filter to minimise the polarising beamsplitter's spectral dispersion property. Four exposures were obtained with the corresponding half-wave plate in positions 0, 45, 22.5 and 67.5 degrees. Each exposure produced both the $e$ and $o$ beam on the detector, each HWP with a field of view of $4\times8$ arc minutes. Bias subtraction proceeded in the usual fashion using the standard SALT data reduction tools and flat fielding. The polarised and unpolarised standard stars Vela1 95 and WD 0310-688 were observed during the same month with the same filter to set the HWP's zero-point and to measure any instrumental polarisation. Aperture photometry was performed on all the sufficiently bright point sources in the science $e$ and $o$ images, and the corresponding linear polarisation was calculated from the extracted fluxes.

\subsection{Spectroscopy}

\css\ was observed spectroscopically at 12 epochs spanning phases between 9.4 and 106.0 days from the start of the outburst. The observations were carried out with five different instruments: ALFOSC at the NOT, MODS \citep{Pogge10} mounted on the twin 8.4-m Large Binocular Telescope (LBT) at Mount Graham International Observatory (AZ, USA), RSS at the SALT, IMACS on the 6.5-m Magellan telescope, and Optical System for Imaging and low-Intermediate-Resolution Integrated Spectroscopy (OSIRIS) at the 10.4-m Gran Telescopio Canarias (GTC). The log of spectroscopic observations of \css\ is presented in Table~\ref{tspectra}. 

The spectra were reduced using standard \textsc{iraf} routines (bias subtraction, flat-field correction, 1D extraction, and wavelength calibration) and custom pipelines (e.g. \textsc{PySALT}, \textsc{foscgui\footnote{\url{https://sngroup.oapd.inaf.it}}}).  The flux calibration was performed using spectra of standard stars obtained during the same night. All spectra are available via the WISeREP\footnote{\href{https://www.wiserep.org/}{https://www.wiserep.org/}} repository \citep{Yaron12}.  
All data (photometry and spectra) are available on Zenodo (\url{https://doi.org/10.5281/zenodo.13844162}).

\begin{table*}
\renewcommand{\thetable}{A\arabic{table}}
\setcounter{table}{0}
\centering
\scriptsize
\caption{Optical photometry of \css$^{\star}$.}
\label{tphoto}
\setlength{\tabcolsep}{3pt}
\begin{tabular}[t]{cccccccccccc}
\hline
\hline

UT date  &   JD        & Phase   &      $B$	       &      $V$       &      $g$       &	 $r$	     &	    $i$	      &   $z$         &  Telescope/Instrument$^{\ddagger}$   \\
         &        & (days)$^{\ast}$ &    (mag)     &     (mag)      &     (mag)      &     (mag)      &     (mag)      &    (mag)     &   &         \\
\hline
\hline                                              
20161006 & 2457667.78 & \nodata &	\nodata        & $<17.54^{**}$  & \nodata        & \nodata        & \nodata        & \nodata        & ASAS-SN    \\
20161010 & 2457671.70 & 1.72    &	\nodata        & $16.52\pm0.12$ & \nodata        & \nodata        & \nodata        & \nodata        & ASAS-SN    \\
20161012 & 2457673.85 & 3.80    &	\nodata        & $15.47\pm0.06$ & \nodata        & \nodata        & \nodata        & \nodata        & ASAS-SN    \\
20161013 & 2457674.95 & 4.87    &	\nodata        & $15.68\pm0.07$ & \nodata        & \nodata        & \nodata        & \nodata        & ASAS-SN    \\
20161015 & 2457676.94 & 6.79    &	\nodata        & $16.45\pm0.17$ & \nodata        & \nodata        & \nodata        & \nodata        & ASAS-SN    \\
20161018 & 2457679.59 & 9.36    &	\nodata        & \nodata        & $16.75\pm0.08$ & $16.89\pm0.11$ & $17.21\pm0.10$ & $17.14\pm0.09$ & LT/IO:O    \\
20161018 & 2457679.63 & 9.39    &	$17.02\pm0.04$ & $16.94\pm0.05$ & $16.96\pm0.03$ & $16.95\pm0.09$ & $17.01\pm0.09$ & $17.10\pm0.08$ & NOT/ALFOSC \\
20161018 & 2457680.11 & 9.86    &	$17.19\pm0.08$ & \nodata        & \nodata        & \nodata        & $17.31\pm0.10$ & \nodata        & LCOGT      \\
20161020 & 2457681.87 & 11.56   &	$17.56\pm0.13$ & $17.76\pm0.17$ & \nodata        & $17.99\pm0.21$ & $17.66\pm0.25$ & \nodata        & LCOGT      \\
20161021 & 2457682.57 & 12.24   &	\nodata        & \nodata        & $17.65\pm0.13$ & $17.80\pm0.06$ & $18.06\pm0.06$ & $18.20\pm0.09$ & LT/IO:O    \\
20161022 & 2457683.57 & 13.21   &	\nodata        & \nodata        & $17.89\pm0.14$ & $18.00\pm0.06$ & $18.22\pm0.10$ & $18.04\pm0.18$ & LT/IO:O    \\
20161023 & 2457684.57 & 14.17   &	\nodata        & \nodata        & $18.12\pm0.13$ & $18.20\pm0.04$ & $18.50\pm0.07$ & $18.40\pm0.10$ & LT/IO:O    \\
20161026 & 2457688.39 & 17.87   &	$19.01\pm0.14$ & $18.82\pm0.16$ & \nodata        & \nodata        & \nodata 	   & \nodata        & LCOGT      \\
20161029 & 2457691.49 & 20.87   &	$19.26\pm0.10$ & $19.29\pm0.16$ & \nodata        & $19.11\pm0.11$ & $19.23\pm0.19$ & \nodata        & LCOGT      \\
20161030 & 2457691.67 & 21.04   &	$19.40\pm0.04$ & $19.47\pm0.08$ & \nodata        & $19.38\pm0.06$ & $19.17\pm0.11$ & $19.33\pm0.18$ & NOT/ALFOSC \\
20161101 & 2457693.61 & 22.92   &	\nodata        & \nodata        & $19.58\pm0.15$ & $19.57\pm0.11$ & $19.82\pm0.13$ & $19.61\pm0.18$ & LT/IO:O    \\
20161102 & 2457694.52 & 23.80   &	\nodata        & \nodata        & $19.62\pm0.14$ & $19.65\pm0.06$ & $19.85\pm0.09$ & $19.65\pm0.14$ & LT/IO:O    \\
20161102 & 2457695.08 & 24.34   &	$19.94\pm0.15$ & $19.63\pm0.17$ & \nodata        & $19.81\pm0.18$ & $20.03\pm0.37$ & \nodata        & LCOGT      \\
20161105 & 2457697.50 & 26.68   &	$19.87\pm0.16$ & \nodata        & \nodata        & $20.22\pm0.26$ & \nodata        & \nodata        & LCOGT      \\
20161110 & 2457702.55 & 31.57   &	\nodata        & \nodata        & $20.55\pm0.21$ & $20.71\pm0.16$ & $21.04\pm0.26$ & $20.01\pm0.13$ & LT/IO:O    \\
20161110 & 2457702.76 & 31.77   &	$20.80\pm0.14$ & $20.73\pm0.17$ & \nodata        & $20.75\pm0.12$ & $20.92\pm0.39$ & \nodata        & LCOGT      \\
20161118 & 2457710.88 & 39.63   &	$22.39\pm0.81$ & $21.49\pm0.34$ & \nodata        & $21.93\pm0.54$ & $22.17\pm1.02$ & \nodata        & MDM        \\
20161119 & 2457711.51 & 40.24   &	\nodata        & \nodata        & $22.21\pm0.38$ & $21.43\pm0.18$ & $22.06\pm0.42$ & $21.98\pm0.56$ & LT/IO:O    \\
20161120 & 2457713.49 & 42.15   &	\nodata        & \nodata        & $22.45\pm0.24$ & $21.87\pm0.10$ & $21.98\pm0.35$ & $21.19\pm0.20$ & LT/IO:O    \\
20161129 & 2457721.64 & 50.04   &	\nodata        & \nodata        & \nodata        & $22.55\pm0.48$ & \nodata        & \nodata        & LT/IO:O    \\
20161205 & 2457727.59 & 55.80   &	$23.50\pm0.86$ & $22.42\pm0.49$ & $24.00\pm0.67$ & $23.19\pm0.39$ & \nodata        & \nodata        & NOT/ALFOSC \\
20161210 & 2457732.57 & 60.61   &	\nodata        & $22.88\pm0.85$ & $24.99\pm2.80$ & $22.94\pm0.48$ & $23.04\pm0.97$ & \nodata        & NOT/ALFOSC \\
20161224 & 2457746.67 & 74.26   &	\nodata        & \nodata        & \nodata        & $23.97\pm0.56$ & \nodata        & \nodata	    & Mag/LDSS-3 \\
20161227 & 2457749.53 & 77.02   &	\nodata        & \nodata        & \nodata        & $22.94\pm0.27$ & \nodata        & $22.07\pm0.38$ & LT/IO:O    \\
20161227 & 2457750.45 & 77.91   &	\nodata        & \nodata        & $22.24\pm0.95$ & \nodata        & $23.04\pm0.90$ & \nodata        & LT/IO:O    \\
\hline
\hline
\multicolumn{11}{c}{\textbf{Host}}\\
\hline
20170721 & 2457955.91 &	276.69  & \nodata          &  \nodata       & $21.67\pm0.04$ & $21.26\pm0.03$ & $20.97\pm0.11$ & $20.74\pm0.08$ & Mag/LDSS-3 \\	
20170919 & 2457984.87 &	304.71  & $22.17\pm0.23$   & $21.64\pm0.22$ &  \nodata       & \nodata        & \nodata        & \nodata        & Mag/IMACS \\	
20170922 & 2458018.65 & 337.39  & $22.03\pm0.14$   & $21.44\pm0.12$ & $21.59\pm0.26$ & $21.17\pm0.18$ & $20.84\pm0.27$ & $20.57\pm0.24$ & NOT/ALFOSC\\
\hline
\end{tabular}
\begin{list}{}{}
\item \textbf{Notes:} \\
$^{\star}$ All reported magnitudes are host-subtracted.\\
\textbf{$^{**}$ 3-sigma upper limit.} \\
$^{\ast}$ Rest-frame phase in days from the start of the outburst, JD $=2457669.92\pm2.00$.\\ 
$^{\ddagger}$ Telescope code: LCOGT: Las Cumbres Observatory Global Network; LT: 2.0-m Liverpool Telescope; 
Mag/IMACS: Inamori Magellan Areal Camera and Spectrograph on Magellan; Mag/LDSS-3: The Low Dispersion Survey Spectrograph on Magellan;
MDM: Hiltner 2.4m Telescope in the MDM observatory; NOT/ALFOSC: Alhambra Faint Object Spectrograph and Camera on the Nordic Optical Telescope (NOT). \\
$BV$ photometry is in the Vega system; $griz$ photometry is in the AB system.
\end{list}
\end{table*}

 % Table 1

\begin{table}
\renewcommand{\thetable}{A\arabic{table}}
\setcounter{table}{1}
\small
\centering
\caption{$BVRI$ photometry obtained with DFOT in the Vega system.}
\label{tphotodfot}
\begin{tabular}[t]{ccccccc}
\hline
\hline
UT date  &     JD     & Phase           &      $B$	     &      $V$       &       $R$      &     $I$        \\
         &            & (days)$^{\ast}$ &      (mag)     &     (mag)      &    (mag)       &    (mag)       \\
\hline         
\hline           
20161022 & 2457684.29 &     13.90       & $18.33\pm0.07$ & $18.09\pm0.05$ &  \nodata       & $17.81\pm0.11$ \\
20161023 & 2457685.43 &     15.00       & $18.55\pm0.06$ & $18.33\pm0.04$ & $18.23\pm0.04$ & $18.02\pm0.11$ \\
20161102 & 2457695.48 &     24.73       & $19.87\pm0.11$ &    \nodata     &     --         &        --      \\
20161108 & 2457701.22 &     30.28       &   \nodata      &    \nodata     &  \nodata       & $19.34\pm0.22$ \\
\hline
\end{tabular}
\begin{list}{}{}
\item \textbf{Notes:} 
\item $^{\ast}$ Rest-frame phase in days from the start of the outburst, JD = $2457669.92\pm2.00$
\end{list}
\end{table}
 % Table 2

\begin{table}
\renewcommand{\thetable}{A\arabic{table}}
\setcounter{table}{2}
\centering
\caption{UV photometry obtained with Swift in the AB system$^{\star}$.}
\label{tphotoswift}
\begin{tabular}[t]{ccccccccc}
\hline
\hline
UT date  &     JD     & Phase           &      UVW1      &      UVM2      &     UVW2       \\
         &            & (days)$^{\ast}$ &      (mag)     &     (mag)      &    (mag)       \\
\hline                                
\hline                                
20161103 & 2457695.90 &	   25.14        & $20.62\pm0.24$ & $20.55\pm0.22$ & $20.60\pm0.19$ \\
20161108 & 2457701.10 &	   30.17        & $21.00\pm0.24$ & $21.18\pm0.22$ & $21.31\pm0.21$ \\
20161111 & 2457704.09 &	   33.06        & $22.44\pm0.74$ & $21.46\pm0.31$ & $21.69\pm0.31$ \\
\hline
\end{tabular}
\begin{list}{}{}
\item \textbf{Notes:} 
\item $^{\star}$ All reported magnitudes are host-subtracted.\\
\item $^{\ast}$ Rest-frame phase in days from the start of the outburst, JD = $2457669.92\pm2.00$.
\end{list}
\end{table}
 % Table 3

\begin{table}
\centering
\renewcommand{\thetable}{A\arabic{table}}
\setcounter{table}{3}
\caption{$H$ photometry in Vega system.}
\label{tphotoltH}
\begin{tabular}[t]{ccccccc}
\hline
\hline
UT date   &	   JD	   & Phase           &  Magnitude    \\
          &            & (days)$^{\ast}$ &   (mag)       \\
\hline                                                           
\hline                                                           
20161018  & 2457679.60 &      9.37      & $16.51\pm0.10$ \\
20161021  & 2457682.58 &      12.24     & $16.99\pm0.08$ \\
20161022  & 2457683.57 &      13.21     & $17.21\pm0.12$ \\
20161023  & 2457684.57 &      14.17     & $17.42\pm0.16$ \\
20161029  & 2457690.61 &      20.02     & $17.60\pm0.18$ \\
20161030  & 2457691.59 &      20.97     & $17.60\pm0.19$ \\
20161031  & 2457692.55 &      21.89     & $17.70\pm0.23$ \\
\hline
\end{tabular}
\begin{list}{}{}
\item \textbf{Notes:} 
Rest-frame phase in days from the start of the outburst, JD $=2457669.92\pm2.00$. \end{list}
\end{table} % Table 4

\begin{table}
\centering
\renewcommand{\thetable}{A\arabic{table}}
\setcounter{table}{4}
\caption{ATLAS AB optical photometry.}
\label{tphotoatlas}
\begin{tabular}[t]{ccccccc}
\hline
\hline
UT date   &	   JD	   & Phase           &  Band  &   Magnitude    \\
          &            & (days)$^{\ast}$ &        &    (mag)       \\
\hline                                                           
\hline                                                           
20161006  & 2457668.14 & \nodata& $c$ &  $<19.59$      \\
20161010  & 2457672.12 & 2.13   & $o$ & $16.36\pm0.02$ \\
20161015  & 2457677.08 & 6.93   & $o$ & $16.38\pm0.11$ \\
20161018  & 2457681.11 & 10.82  & $o$ & $17.39\pm0.05$ \\
20161027  & 2457689.10 & 18.56  & $c$ & $18.85\pm0.11$ \\ 
20161104  & 2457697.08 & 26.28  & $c$ & $20.22\pm0.26$ \\
20161107  & 2457700.05 & 29.15  & $c$ & $20.16\pm0.37$ \\ 
20161108  & 2457701.06 & 30.13  & $o$ & $20.27\pm0.64$ \\
20161112  & 2457705.01 & 33.95  & $o$ & $<21.51$       \\
20161120  & 2457713.03 & 41.71  & $o$ & $<21.40$       \\
20161124  & 2457717.04 & 45.59  & $c$ & $<20.51$       \\
\hline
\end{tabular}
\begin{list}{}{}
\item \textbf{Notes:} 
Rest-frame phase in days from the start of the outburst, JD $=2457669.92\pm2.00$. \end{list}
\end{table} % Table 5

\begin{table*}
\renewcommand{\thetable}{A\arabic{table}}
\setcounter{table}{5}
\centering
\caption{Spectroscopic observations of \css}
\label{tspectra}
\begin{tabular}[t]{cccccccccc}
\hline
\hline
UT date  &	   JD     &       Phase    	&  Range       &  Telescope   & Grism/Grating  \\
         &            & (days)$^{\ast}$ &  (\AA)       & +Instrument  &                \\
\hline
\hline
20161018 & 2457679.64 &   9.40       & $3280-9320$  &  NOT+ALFOSC  &    Grism\#4    \\
20161019 & 2457680.67 &   10.40      & $3290-9380$  &  NOT+ALFOSC  &    Grism\#4    \\
20161020 & 2457681.59 &   11.29      & $3290-9370$  &  NOT+ALFOSC  &    Grism\#4    \\
20161030 & 2457691.65 &   21.02      & $3280-9320$  &  NOT+ALFOSC  &    Grism\#4    \\
20161031 & 2457692.66 &   22.00      & $3290-9370$  &  NOT+ALFOSC  &    Grism\#4    \\
20161104 & 2457697.44 &   26.62      & $3700-8300$  &   SALT+RSS   &    PG0300      \\
20161105 & 2457697.73 &   26.91      & $4030-9050$  &   Mag+IMACS  &    Gri-300-17.5\\
20161106 & 2457698.78 &   27.92      & $4030-9050$  &   Mag+IMACS  &    Gri-300-17.5\\
20161119 & 2457711.88 &   40.60      & $2980-9670$  &   LBT+MODS   &    G400L/G670L \\
20161124 & 2457716.55 &   45.11      & $3520-7620$  &   GTC+OSIRIS &  R1000B/R1000R \\
20161206 & 2457729.58 &   57.72      & $5100-10400$ &   GTC+OSIRIS &    R1000R      \\
20170125 & 2457779.48 &   106.00     & $4520-7615$  &   GTC+OSIRIS &    R1000B      \\
\hline 
\end{tabular}
\begin{list}{}{}
\item \textbf{NOTES:}
\item $^{\ast}$ Rest-frame phase in days from the start of the outburst, JD$=2457669.92\pm2$.
\item \textbf{Telescope code:} \textbf{GTC:} Gran Telescopio Canarias; \textbf{LBT:} Large Binocular Telescope; \textbf{Mag:} 6.5-m Magellan telescope; \textbf{NOT:} Nordic Optical Telescope; \textbf{SALT:}  South African Large Telescope. 
\end{list}
\end{table*}

 % Table 6

\clearpage
\section{Host galaxy analysis}
\label{ap:host}

Using the host galaxy GTC/OSIRIS spectrum (JD$=2457779.48$) covering the wavelength range $3600-7900$ \AA, and $BgVriz$ photometry (JD$=2457955.91$ and 2457984.87), we characterised \css's host properties. We use \textsc{prospector} \citep{Johnson21-prosp}, a versatile stellar population fitting tool that uses Monte Carlo sampling of the posterior distributions with \textsc{emcee} \citep{Foreman-Mackey13}. The photometry and spectroscopy are simultaneously fit, ensuring a proper spectrum calibration by optimizing the parameters of a polynomial that multiplies the model spectrum to match the observed spectrum at each iteration in the fitting process. Therefore, the spectral continuum does not influence the inferred physical stellar parameters. A free $\sigma_v$ parameter also ensures that the stellar model spectra are smoothed to the same resolution as the observed spectrum. We use the \textsc{miles} stellar libraries \citep{Sanchez-Blazquez06} as provided by the Flexible Stellar Population Synthesis code \textsc{fsps} \citep{Conroy09, Foreman-Mackey14}, and a non-parametric star-formation history consisting of eight age bins. Seven age bins have a free amplitude parameter, while the eighth is constrained by the total stellar mass formed. A single stellar metallicity is inferred, and a dust screen model \citep{Kriek13} is assumed to affect all stars and with two free parameters for the dust optical depth $\tau_V$ and the slope of the attenuation curve. The young stars in star-forming regions ($<10^7$yr) are also affected by an additional dust component parameterized by the prescription of \citet{Blitz80}, including a free gas dust fraction parameter. We also simultaneously fit the nebular part of the spectrum, which has two additional parameters, the gas ionisation parameter and the gas-phase metallicity. Thus, we have 15 free parameters inferred through a Monte Carlo sampling with 200 walkers and 2000 iterations. The best-fit models compared to the data, the corner plots of some parameters and the star-formation history are shown in Figure~\ref{fig:galcorner}. The 16th, 50th and 84th percentiles of all parameters are shown in Table~\ref{tgalpar}.

\begin{figure}
\setcounter{figure}{0}
\renewcommand{\thefigure}{B\arabic{figure}}
\renewcommand{\theHfigure}{B\arabic{figure}}
\centering
\includegraphics[width=0.45\textwidth]{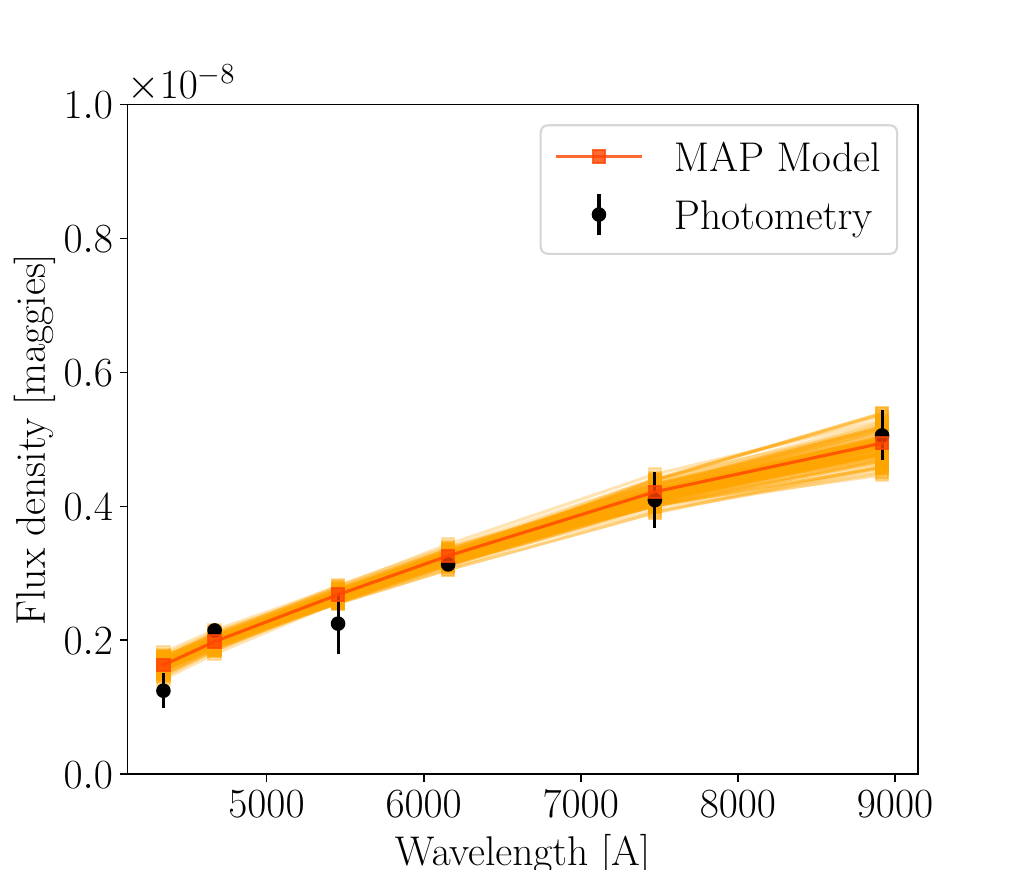}
\includegraphics[width=0.45\textwidth]{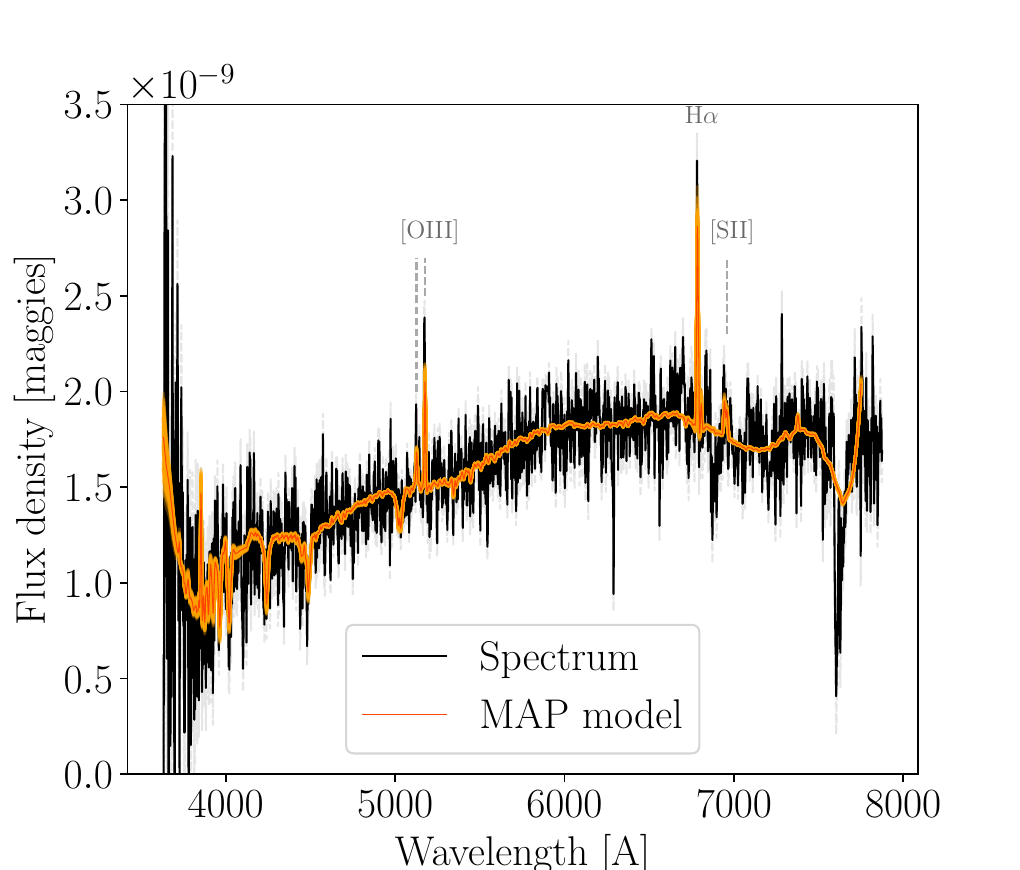}
\includegraphics[width=0.68\textwidth]{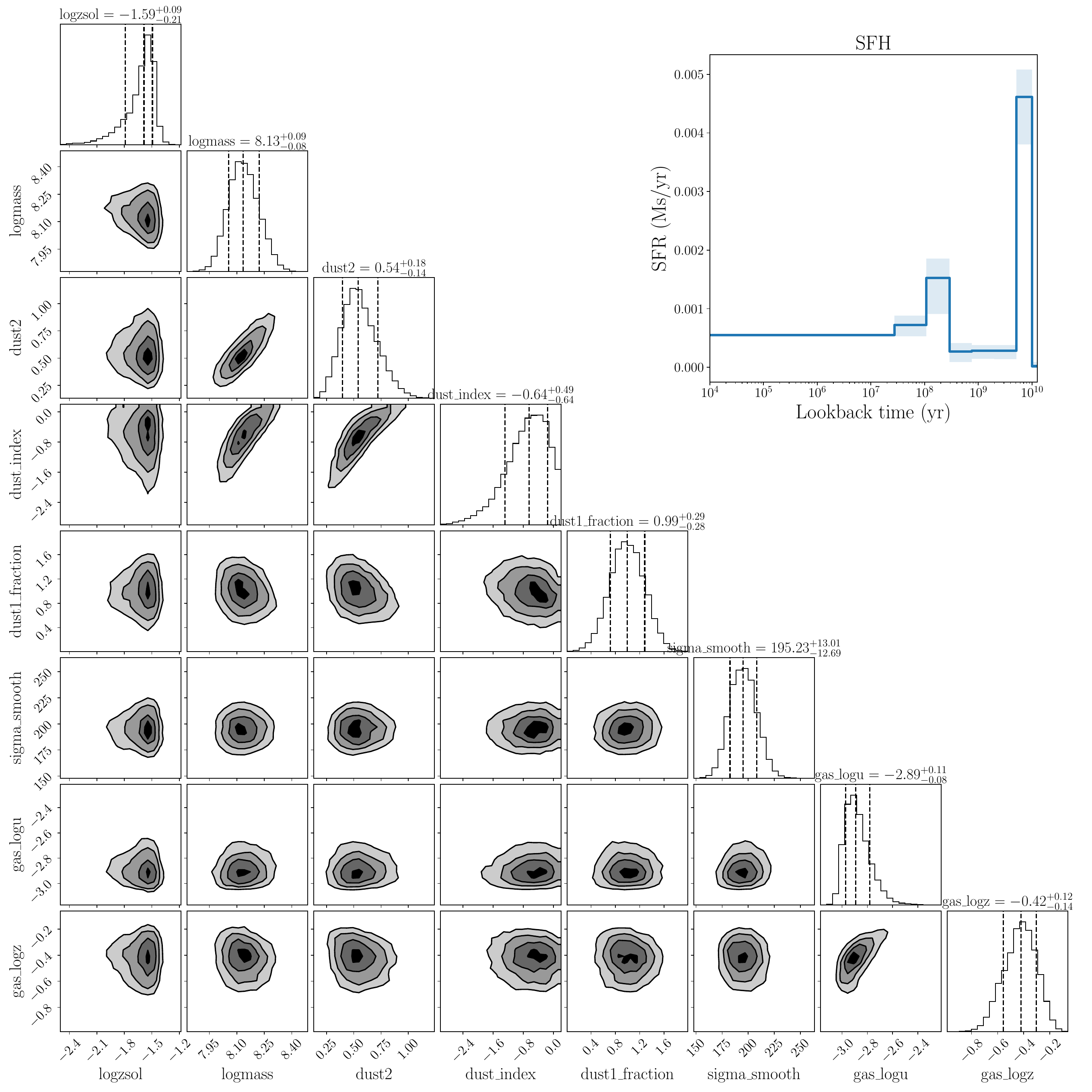}
\caption{Fit results from \textsc{prospector} obtained for the photometric SED (upper left) and spectroscopic SED (upper right) of \css's galaxy: red lines and point show the best model (Maximum A Posteriori or MAP) while the orange lines show 100 random samples of the posterior. The corner plot shows the sampling of the posterior probability distribution for the parameters obtained using \textsc{prospector} (excluding the star-formation history amplitudes). The middle right plot shows the non-parametric star-formation history inferred through eight age bins. The median and 16\% and 84\% percentiles of \css's host parameters are shown in Table~\ref{tgalpar}.}
\label{fig:galcorner}
\end{figure}

The velocity dispersion parameter we obtained, $\sigma_v=195.23^{+13.01}_{-12.69}$ \kms\ is dominated by the broadening needed to go from the spectral resolution of the templates to that of the instrument (295 \kms). Higher-resolution spectra are necessary for an accurate velocity dispersion that could help constrain the BH mass independently through the $M-\sigma$ relation \citep{Ferrarese00}. 

We also derived the gas phase metallicity using the emission lines from the H~II region near \css. By measuring the fluxes of \ha, \hb, [O~III] $\lambda5007$ and [N~II] $\lambda6583$, and applying the O3N2 and N2 diagnostic methods from \citet{Marino13}, and the diagnostic from \citet{Dopita16}, we obtained an oxygen abundance of 12 + log(O/H)$=8.06\pm0.05$ dex, 12 + log(O/H)$=8.22\pm0.12$ dex and 12 + log(O/H)$=8.03\pm0.04$ dex, respectively. These estimates are consistent and suggest a low metallicity (0.23 \Zsun), which also agrees quite well with the gas-phase metallicity found with \textsc{prospector} (0.379$^{+0.077}_{-0.118}$ \Zsun). The difference between the much lower stellar metallicity (0.026$^{+0.007}_{-0.010}$ \Zsun) and the gas-phase metallicity could arise from two different star-formation episodes in the host galaxy: a very old primordial burst that created most of the old stars with extremely low metallicity, and a low but non-null episode of recent star-formation that explains the nebular line presence and the higher gas-phase metallicity. This interpretation agrees well with the estimated star-formation history that shows two major peaks: one in the oldest bin ($>10^{10}$yr) and a more recent one spanning the youngest four bins ($<10^{8.5}$yr). Similar results have been previously found for other dwarf galaxies \citep{Gallazzi05, Lian18, Fraser-McKelvie22}. 

A detailed analysis of WISEA J045834.37-081804.4 was presented in 2020 by \citetalias{Coppejans20}. Using an optical spectrum and photometry they found a current stellar age of $(0.6-4)$\,Gyr, a stellar mass of $M_* \approx10^{7}$\,\Msun, a star formation rate of $\mathrm{SFR} = 4\times10^{-3}$ \Msun yr$^{-1}$, and a specific star formation rate of $\mathrm{sSFR}=0.3$\,Gyr$^{-1}$. They pointed out that their estimated stellar mass would indicate a central black hole with an intermediate mass of $\sim10^3$ \Msun, possibly even lower than our estimate. Most of these parameters resemble ours, although the stellar mass differs from our estimations by an order of magnitude. This is mostly due to our addition of $z$-band photometry that further constrains the stellar mass. However, we noticed that if we run \textsc{prospector} with the gas phase and stellar metallicities fixed to the value found from our host spectrum, we find a stellar mass of log(M$_{*}/$\Msun)$=7.02^{+0.17}_{-0.14}$, which is similar to the mass obtained by \citetalias{Coppejans20}. Nevertheless, we favour our initial result, where the gas phase and stellar metallicities are free parameters because they provide very different information on the galaxy evolution \citep{Fraser-McKelvie22}.

\begin{table}
\setcounter{table}{0}
\renewcommand{\thetable}{B\arabic{table}}
\renewcommand{\theHtable}{B\arabic{figure}}
\centering
\caption{Median and 16\% and 84\% percentiles of \css's host parameters obtained with \textsc{prospector}}
\label{tgalpar}
\begin{tabular}[t]{cccc}
\hline
\hline
SFR ($10^{-3}$\Msun/yr) at $0.0<\log(t/yr)<7.5$   & $0.698^{+0.028}_{-0.052}$ \\  
SFR ($10^{-3}$\Msun/yr) at $7.5<\log(t/yr)<8.0$   & $0.489^{+0.129}_{-0.193}$ \\ 
SFR ($10^{-3}$\Msun/yr) at $8.0<\log(t/yr)<8.42$   & $1.833^{+0.632}_{-1.089}$ \\
SFR ($10^{-3}$\Msun/yr) at $8.4<\log(t/yr)<8.8$   & $0.571^{+0.219}_{-0.238}$ \\
SFR ($10^{-3}$\Msun/yr) at $8.8<\log(t/yr)<9.3$   & $0.029^{+0.054}_{-0.022}$ \\
SFR ($10^{-3}$\Msun/yr) at $9.3<\log(t/yr)<9.7$   & $0.003^{+0.003}_{-0.002}$ \\  
SFR ($10^{-3}$\Msun/yr) at $9.7<\log(t/yr)<10.1$   & $0.011^{+0.016}_{-0.007}$ \\  
SFR ($10^{-3}$\Msun/yr) at $10.1<\log(t/yr)$$^{\diamond}$ & $2.651^{+0.634}_{-1.559}$ \\ 
Log(Z/\Zsun)              & $-1.586^{+0.092}_{-0.206}$ \\
Log(M$_{*}/$\Msun)        & $8.135^{+0.087}_{-0.079}$  \\
Log(Age/yr)$^{\ast}$      & $7.678^{+0.690}_{-0.623}$  \\
Recent SFR(\Msun/yr)      & $0.015^{+0.010}_{-0.008}$  \\
Dust ($\tau_V$)           & $0.542^{+0.179}_{-0.144}$  \\
Dust\_index ($n$)         & $-0.644^{+0.487}_{-0.638}$ \\
Gas dust fraction (\%)$^{\dagger}$  & $0.994^{+0.288}_{-0.278}$  \\
Gas ionisation parameter  & $-2.892^{+0.111}_{-0.081}$ \\
Log(Z$_{gas}$/\Zsun)      & $-0.421^{+0.118}_{-0.136}$ \\
$\sigma_v$ (km/s)         & $195.234^{+13.010}_{-12.693}$ \\
\hline
\end{tabular}
\begin{list}{}{}
\item \textbf{NOTES:}
\item $^{\ast}$ Mass-weighted age obtained from the total stellar mass and the star-formation rates bins of the SFH.
\item $^{\dagger}$ Fraction of dust ($\tau_V$) that also affects the young stellar populations ($<10^7$yr).
\item $^{\diamond}$ The last SFR bin is obtained from the other seven bins and the total stellar mass formed. 
\end{list}
\end{table} % Table 7

\section{\ha\ emission profile properties}
\label{ap:ha}

To quantify the evolution of \ha\ in \css, we measured its equivalent width (EW), velocity offset, full width at half-maximum (FWHM), the highest velocity indicated by the bluest part of the line profile and luminosity (Figure~\ref{fig:Hameasurements}). The EW, FWHM, and highest velocity are most affected as the line becomes weaker. The EW evolves from $\sim250$ at 21.0 days to $\sim30$ \AA\ at 57.7 days, while the FWHM velocity drops from $28000$ to $4000$ \kms\ during the same period. The highest velocities of \ha\ and \hb\ (the bluest parts of the profile) decrease from $\sim-33000$ \kms\ at 21.0 days (consistent with the velocities measured for \heii) to $\sim-10000$ \kms\ at 57.7 days, while the line centre always remains blueshifted by more than $-4000$ \kms. The \ha\ luminosity rises for $\sim8$ days, having a peak luminosity of $1.67\times10^{40}$ at $\sim28$ days. Post peak, the luminosity rapidly decreases, and after $\sim40$ days, it has a quasi-constant evolution. 

\begin{figure}
\setcounter{figure}{0}
\renewcommand{\thefigure}{C\arabic{figure}}
\renewcommand{\theHfigure}{C\arabic{figure}}
\centering
\includegraphics[width=0.5\textwidth]{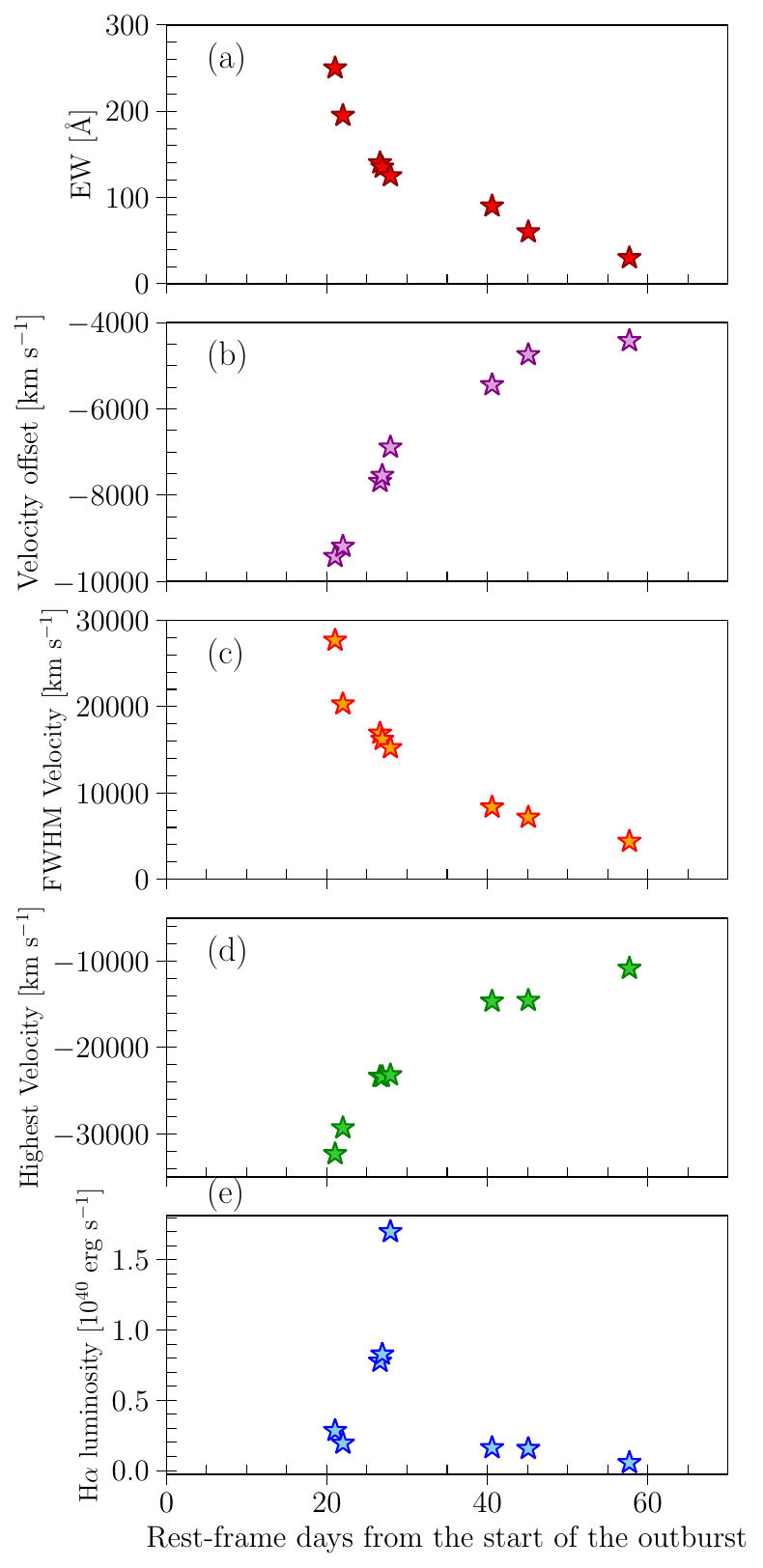}
\caption{Evolution of the EW \textbf{(panel a)}, velocity offset (i.e., blueshift of the line peak from the rest wavelength; \textbf{(panel b)}, FWHM velocity \textbf{(panel c)}, highest velocity \textbf{(panel d)} and luminosity \textbf{(panel e)} of the \ha\ profile from 21.0 to 57.7 days. }
\label{fig:Hameasurements}
\end{figure}

\section{Black hole mass from \css\ light curve}
\label{ap:tde}

Recently, evidence has been found that the TDE properties are related to the BH mass (e.g. \citealt{Blagorodnova17, Wevers17, vanVelzen21, Hammerstein23}). This connection is reinforced by a correlation found between the decay rate of the TDE light curve and the BH mass \citep{Blagorodnova17,  vanVelzen19, vanVelzen20}, where faster-decaying objects have smaller BH masses. Theoretically, the expected decline rate of the post-disruption mass return flow is consistent with a power law decay t$^{-5/3}$ \citep{Rees88, Phinney89}; however, other parameters can affect it \citep{Lodato09} and different indices have been found in observations (from $-0.93$ to $-2.46$ \citealt{vanVelzen21, Hammerstein23}). In fact, recent simulations have shown that the fallback rate from partial TDEs is proportional to a power law decay with $p-$index between $-2$ and $-5$ \citep{Coughlin19, Ryu20}. 

We estimate the $p-$index that best reproduces \css's light curve by fitting a power law decay (L$_{\text{bb}} \propto$ \text{(t/t}${_{0})^{p}}$) to its bolometric light curve. We use Monte Carlo sampling of the posterior distributions with \textsc{emcee} \citep{Foreman-Mackey13} to obtain the free parameters in our model, which are the characteristic time (t$_{0}$) and the power-law index ($p$). We found t$_{0}=2.78^{+0.28}_{-0.26}$ days and $p=-4.09\pm1.30$ (blue line, Figure~\ref{fig:LCfits}). This power-law index is the highest reported to date and is in the range of indexes suggested for a partial TDE \citep{Ryu20}. 

\begin{figure}
\setcounter{figure}{0}
\renewcommand{\thefigure}{D\arabic{figure}}
\renewcommand{\theHfigure}{D\arabic{figure}}
\centering
\includegraphics[width=0.5\textwidth]{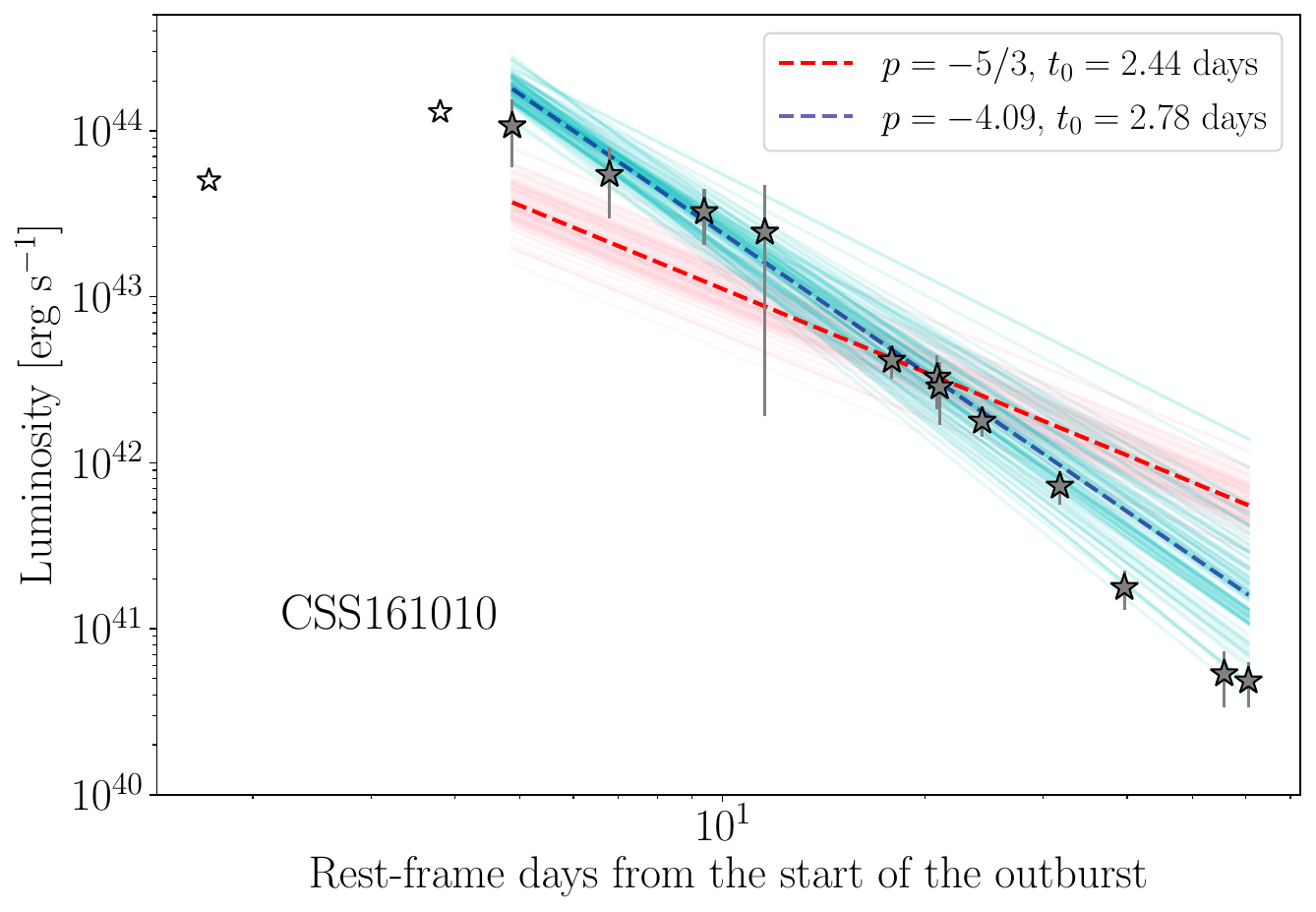}
\caption{Bolometric light curve of \css. The dashed-lines indicated the best-fit for \text{(t/t}$_{0})^{p}$ with $p$ as a free parameter (blue) and $p=-5/3$ (red). Solid lines (red and blue) show 100 random samples of the posterior by using MCMC.}
\label{fig:LCfits}
\end{figure}

To constrain the BH mass of a TDE, we can use the correlation between the fallback timescale (t$_{\text{fb}}$) and the BH mass. As t$_{\text{fb}}$ is comparable to t$_{0}$, we can estimate this parameter by fitting the bolometric light curve of the transient with a power law decay using a fixed power-law index ($p=-5/3$). With the \textsc{emcee} sampler, we fit \css's bolometric light curve and obtain a t$_{0}=2.44^{+0.29}_{-0.27}$ days (red line, Figure~\ref{fig:LCfits}). Extrapolating the correlation between t$_{\text{fb}}$ and BH mass (Figure~\ref{fig:BHM_FBD}) and using t$_{0}=2.44$ days, we find a BH mass of $10^{3.57\pm0.10}$ \Msun. For t$_{0}=2.78^{+0.28}_{-0.26}$ days, we get a BH mass of $10^{3.67\pm0.10}$ \Msun. These BH mass values are in the middle of the range we found using scaling relations (similar to the mass obtained by fitting all galaxies; Figure~\ref{fig:hostgal}; Appendix~\ref{ap:host}. All these findings support \css\ as the fastest declining TDE (candidate) to date with one of the lowest BH masses.

\section{Location of \css\ within its host galaxy}
\label{ap:location}

To investigate the location of \css\ within its host galaxy, we selected pairs of images obtained with ALFOSC on the NOT showing the transient while still bright and the host galaxy after the transient had faded away. For this, we used $B$ and $V$ band images obtained on the nights of 2016 October 29 and 2017 September 21, respectively (Figure~\ref{fig:CSSlocation}). The latter images were obtained with a 900-sec integration per band. The early-time image in each band was aligned to the late-time image using ten isolated and non-saturated stars across the field of view, including shifts in $x$ and $y$ and rotation as the free parameters. The early-time images (with a better seeing) were then convolved to match the late-time images using \textsc{isis 2.2} \citep{Alard98, Alard00} prior to subtraction. The position of the transient was measured from the subtracted images using centroiding. It was compared with the centroid position of the compact host galaxy measured from the late-time images. We find that in R.A., the transient position coincides with the centroid position of the host galaxy within 2$\sigma$. However, in Decl. we find a statistically significant offset of $0\farcs304 \pm 0\farcs032$ 
 in $B$ (10$\sigma$) and $0\farcs383 \pm 0\farcs024$ in $V$ band (16$\sigma$). At the distance of the host galaxy, the $V$ band offset corresponds to a projected distance of 280 pc. In comparison, we measured a FWHM extent of 1.7" (corresponding to 1.2 kpc) for the host galaxy in the late-time $V$ band images with a seeing of FWHM = 1.1". We note that \cite{Coppejans20} found a similar offset between the optical and radio coordinates of the transient and its host galaxy in the $V$ band.

\begin{figure}
\setcounter{figure}{0}
\renewcommand{\thefigure}{E\arabic{figure}}
\renewcommand{\theHfigure}{E\arabic{figure}}
\centering
\includegraphics[width=0.35\textwidth]{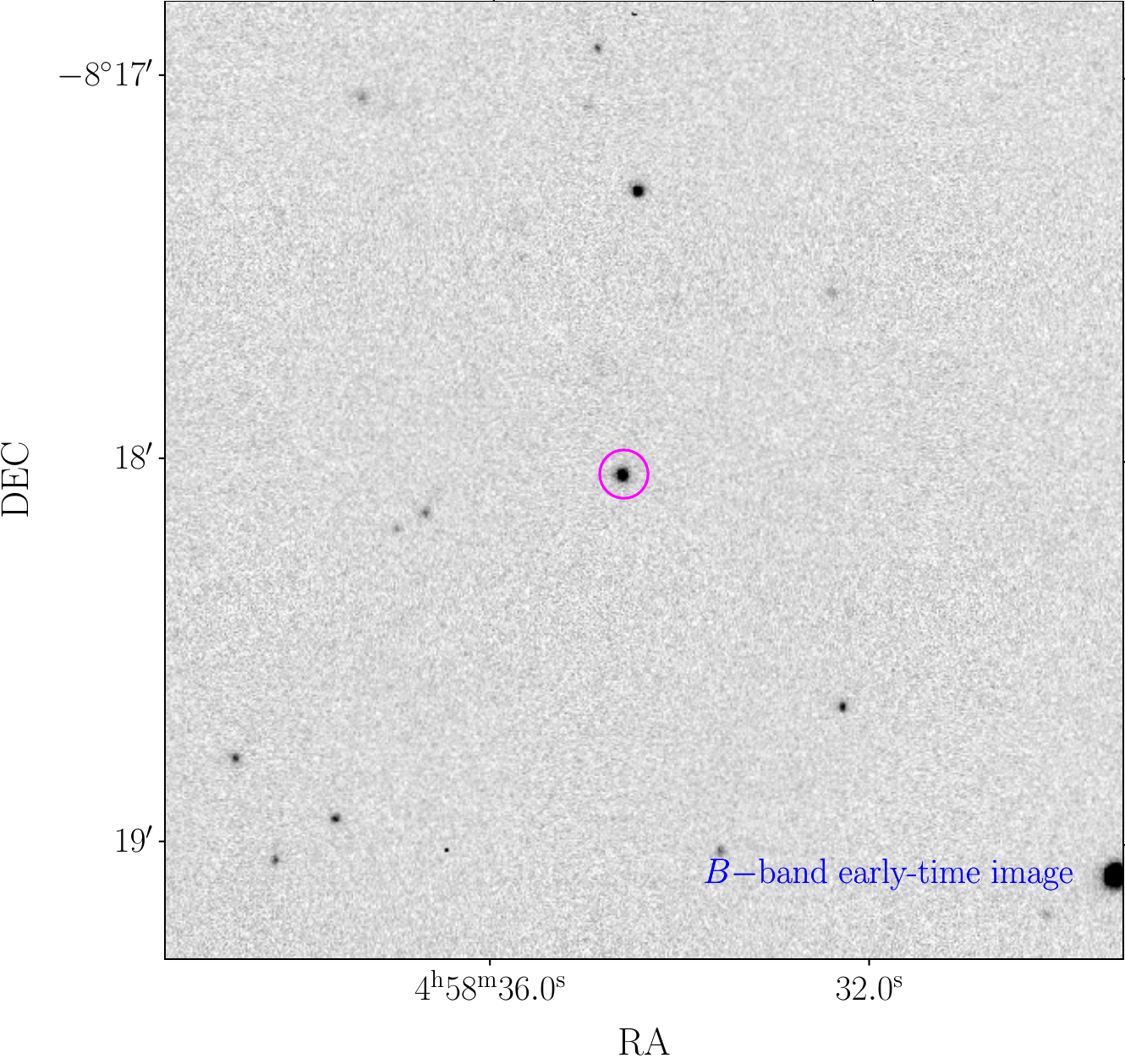}
\includegraphics[width=0.35\textwidth]{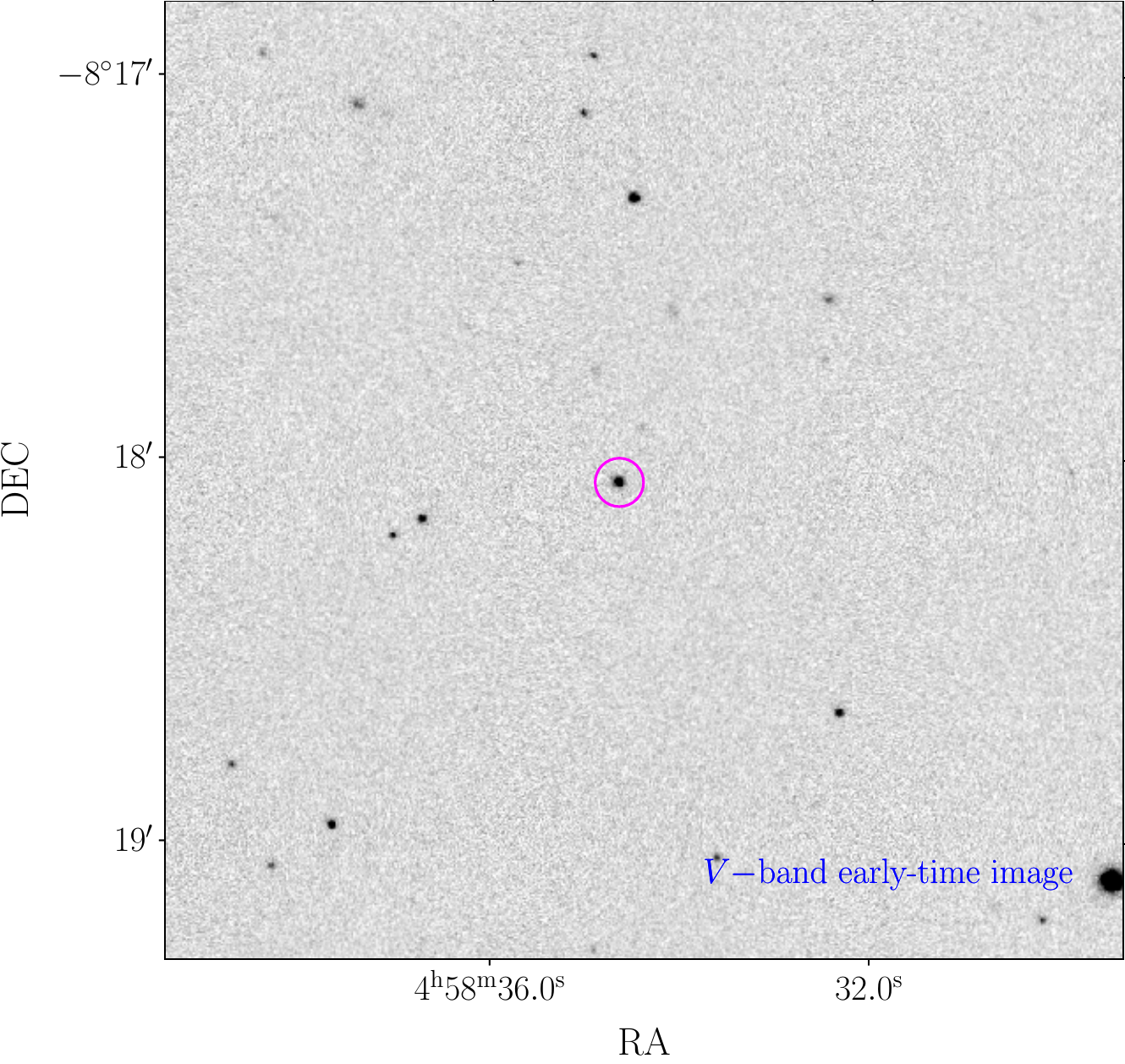}
\includegraphics[width=0.35\textwidth]{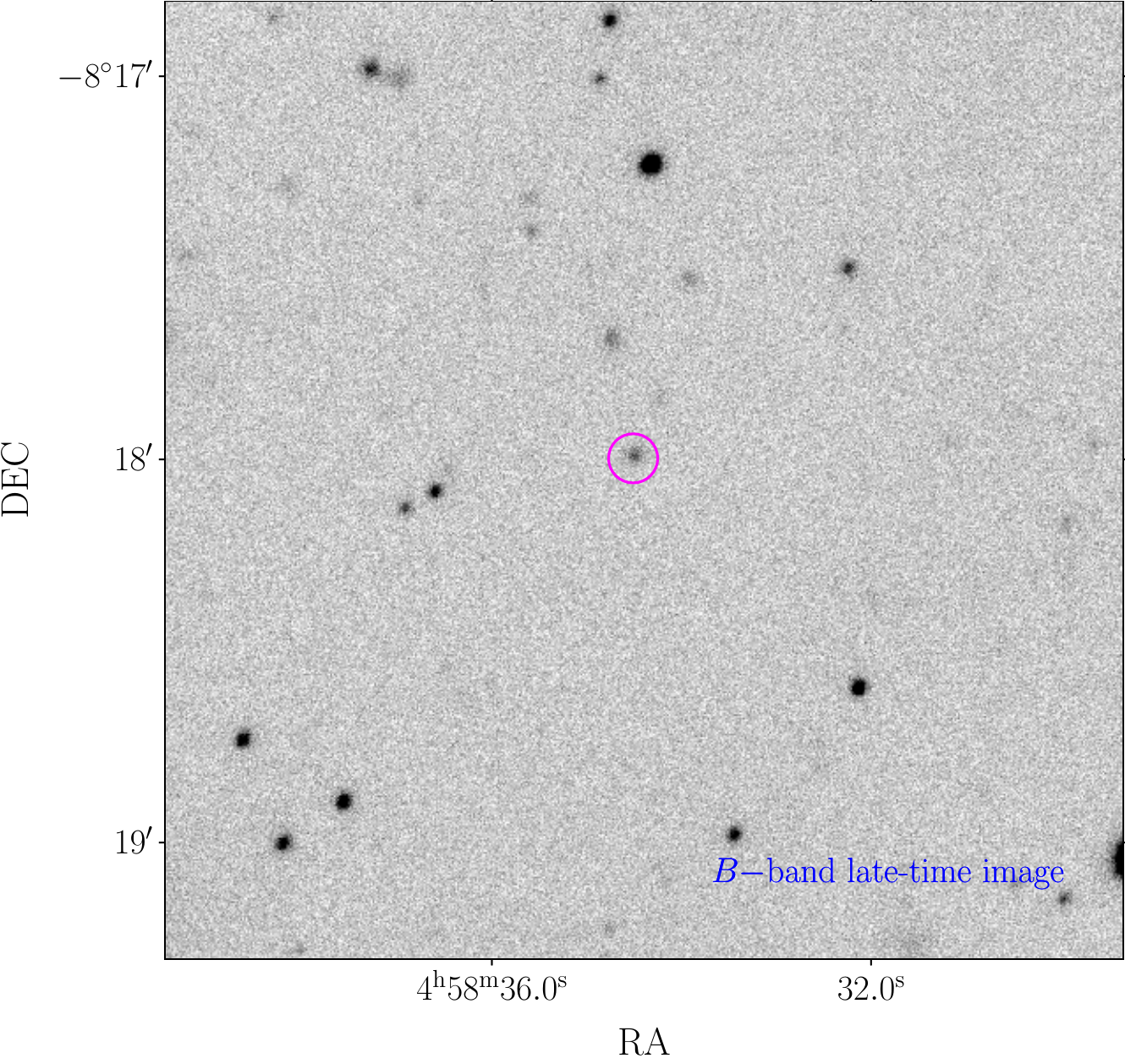}
\includegraphics[width=0.35\textwidth]{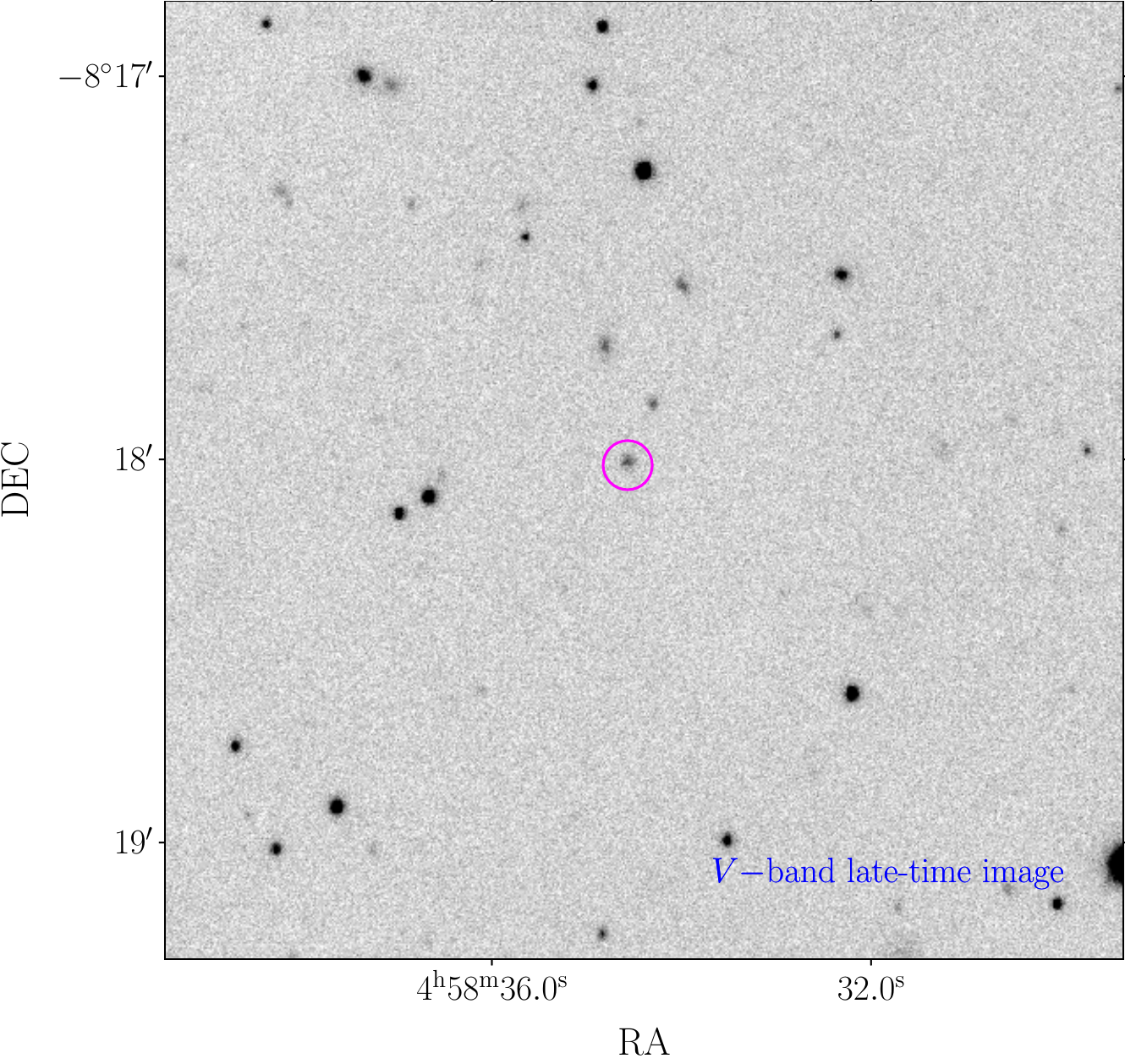}
\includegraphics[width=0.36\textwidth]{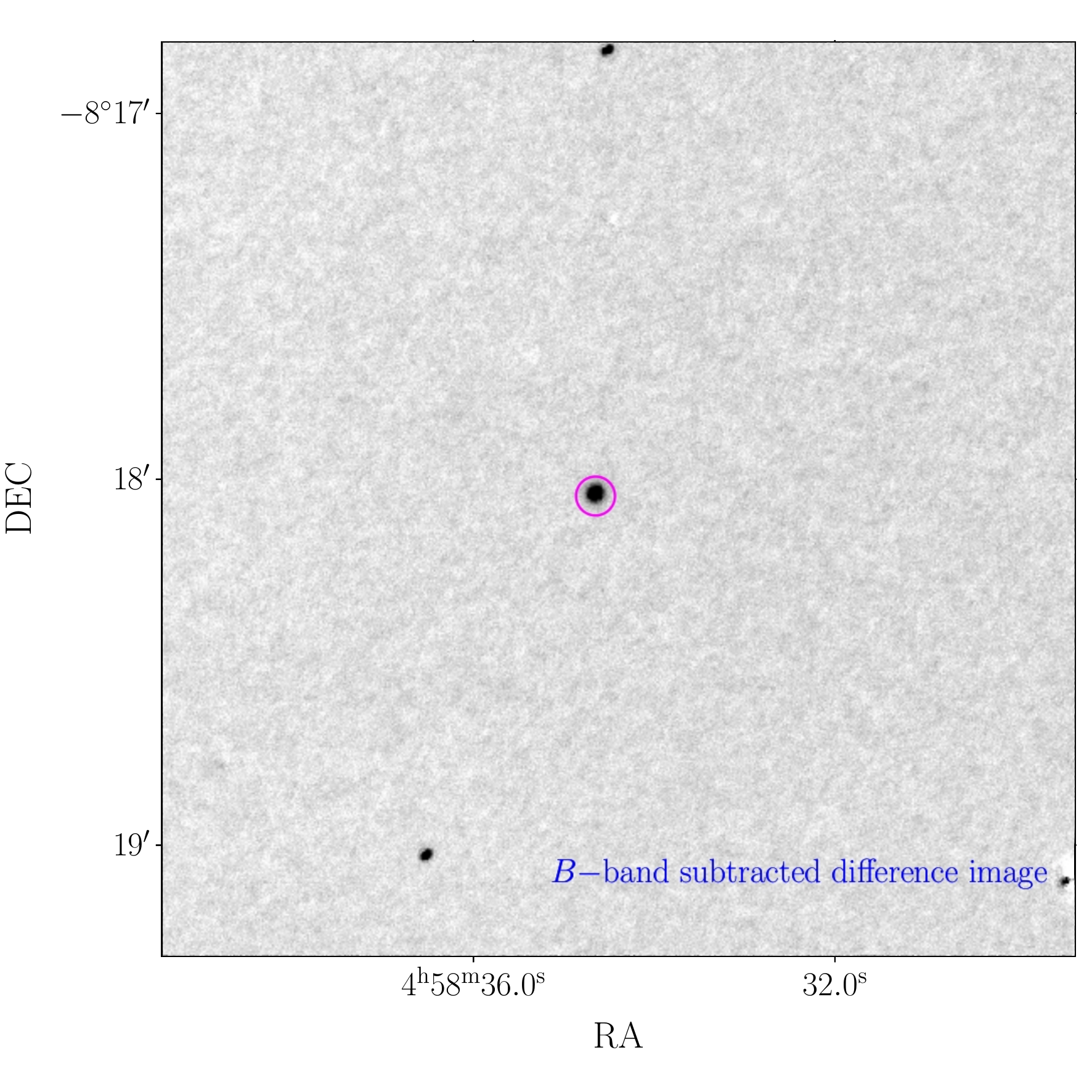}
\includegraphics[width=0.36\textwidth]{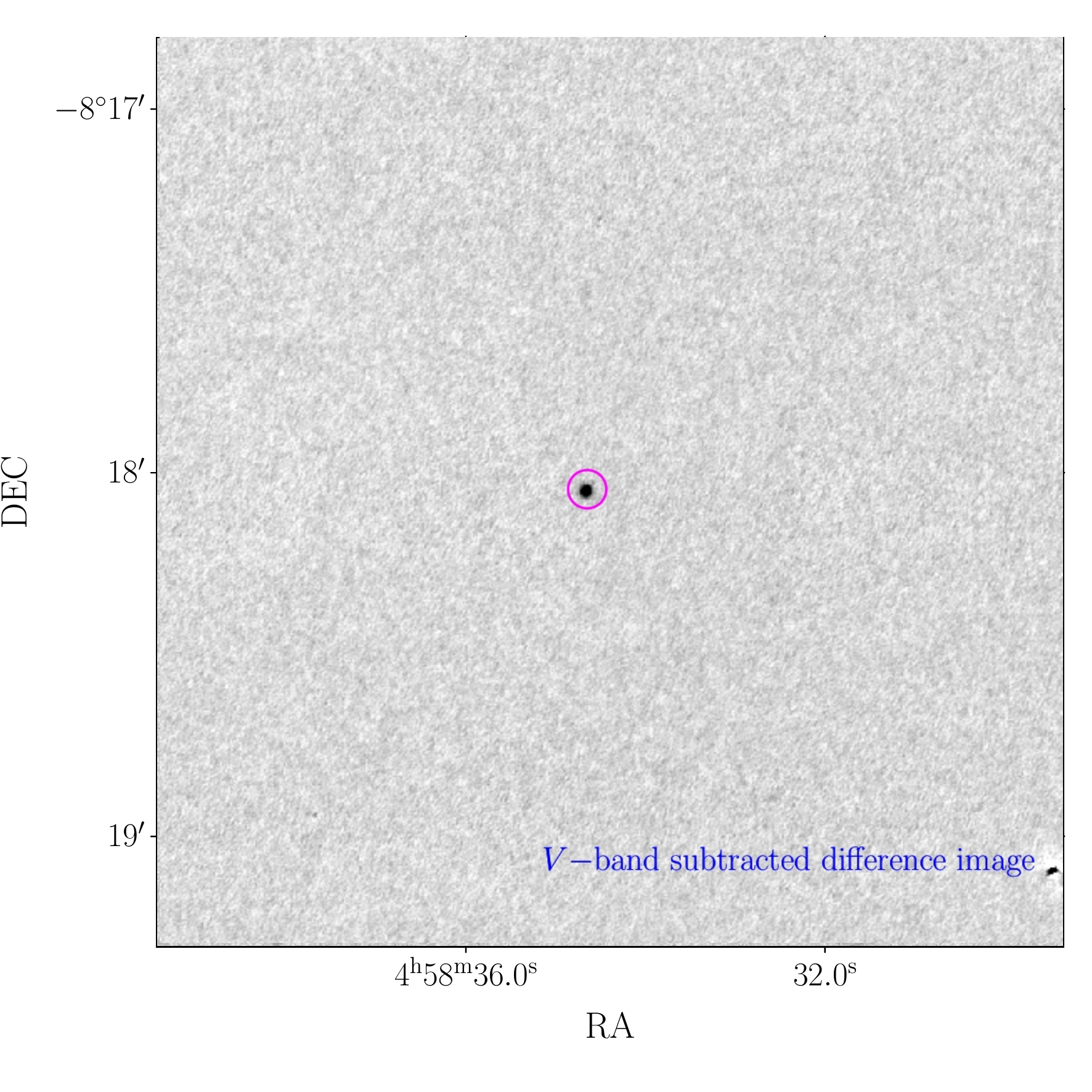}
\caption{Location of CSS161010 within its host galaxy. Images obtained by ALFOSC on the NOT showing the transient (top) and the host galaxy when the transient had faded away (middle). Subtracted difference images (bottom). $B-$band images are shown on the left panels, while V-band images are on the right. }
\label{fig:CSSlocation}
\end{figure}

\clearpage
\section{Other tables}

\begin{table}[!h]
\setcounter{table}{0}
\renewcommand{\thetable}{F\arabic{table}}
\renewcommand{\theHtable}{F\arabic{table}}
\centering
\caption{Detailed properties of the comparison sample}
\label{tcomp}
\setlength{\tabcolsep}{4pt}
\begin{tabular}{ccccccccc}
\hline
\hline
Object     & Redshift & Start of the outburst date (JD) &  References  \\
           &          &                &              \\ 
\hline       
\multicolumn{4}{c}{\textbf{LBOTS}}\\
AT 2018cow & 0.014    &   2458284.84   & \cite{Prentice18, Perley19} \\ 
AT 2020mrf & 0.135    &   2459012.5    & \cite{Yao22}          \\ 
AT 2020xnd & 0.2433   &   2459132.50   & \cite{Perley21, Ho22}           \\ 
\hline       
\multicolumn{4}{c}{\textbf{H-rich SNe}}\\
SN 1979C   & 0.00455  &   2443970.5    &  \cite{Panagia80}       \\ 
SN 2008es  & 0.205    &   2454574.5    &  \cite{Gezari09}        \\ 
\hline       
\multicolumn{4}{c}{\textbf{H-rich TDEs}}\\
AT 2018zr  & 0.071    &   2458156.2   &  \cite{Holoien19, Charalampopoulos22} \\ 
AT 2020neh & 0.062    &   2459018.6   &  \cite{Angus22}                       \\ 
AT 2020wey & 0.0274   &   2459175.5   &  \cite{Charalampopoulos23}            \\ 
\hline
\end{tabular}
\end{table}
 % Table 8


\bibliography{Bibliography}{}
\bibliographystyle{aasjournal}



\end{document}